\documentclass[preprint,pre,aps]{revtex4}
\usepackage{graphics}
\usepackage{amsmath,amssymb}
\usepackage[dvips]{graphicx}
\begin{document}

\title{Vapour-liquid critical parameters of a $2$:$1$ primitive model of ionic fluids confined in disordered porous media}
 \author{Oksana Patsahan, Taras Patsahan, Myroslav Holovko
 }
\affiliation{Institute for Condensed Matter Physics of the National
Academy of Sciences of Ukraine, 1 Svientsitskii St., 79011 Lviv,
Ukraine}

\date{\today}
\begin{abstract}
We study the vapour-liquid  critical parameters of
an ionic fluid confined in  a disordered porous
medium by using the theory which combines the collective variables approach with an extension of the scaled particle theory.
The ionic fluid is described as a two-component charge- and size-asymmetric primitive model, and a porous medium is modelled as
a disordered matrix formed by hard-spheres obstacles.
In  the particular  case of the
fixed valencies $2$:$1$, the coexistence curves and the corresponding critical parameters are calculated for different matrix porosities
as well as for different diameters of matrix and fluid particles. The obtained results show that the general trends of the reduced critical temperature
and the reduced critical density with the microscopic characteristics are similar to the trends obtained in the monovalent case.
At the same time, it is noticed that
an ion charge asymmetry  significantly weakens the  effect of the matrix presence.
\end{abstract}
 \maketitle
 
 \section{Introduction}
This paper is dedicated to the memory of Lesser Blum who passed away last year. His analytical solutions of
the Ornstein-Zernike  integral equation for different fluid models continue to play an important role  in  a modern liquid matter
theory.

When a fluid is confined in a porous medium, its physical properties including the phase behavior  may drastically change
\cite{Gelb,Chandra}.
A fundamental understanding  of confinement effects is of great importance in many industrial applications.
The prediction of the phase equilibria in  ionic systems confined in disordered porous media
represents one of the most important problems in physical chemistry and chemical engineering.

The  phase diagrams and criticality in the bulk ionic fluids
have been intensively  studied experimentally as well as by  using theoretical  and computer simulation methods (see
\cite{PatMryg12} and references cited therein). The main  attention has been paid to the so-called Coulombic systems in
which the phase separation is primarily driven by Coulomb interactions \cite{Wein01}. Molten salts, electrolytes in solvents
of low dielectric constant and  ionic liquids are examples of such Coulombic systems.
The  most frequently used theoretical model for the systems dominated by Coulomb
interactions  is a two-component primitive model (PM).
In this model, the ionic fluid is
described as an electroneutral mixture of charged hard spheres with diameters $\sigma_{+}$ and $\sigma_{-}$
immersed in a structureless dielectric continuum.
The specific feature of the PM is the existence of the vapour-liquid-like phase transition at
low reduced temperature and at low reduced density.
The phase behavior of such systems
confined in microporous materials has received little attention.

A theoretical model to study the systems in disordered confinement mainly relies on the so-called partly-quenched
mixture, where  the immobilized particles of the quenched species constitute the matrix and the particles of
the mobile (annealed) species represent the adsorbate  \cite {Madden88}.  In this case, statistical-mechanical averages
used for calculations of
thermodynamic properties of a fluid
distributed inside a disordered microporous matrix involve taking a double average:
first over the configurational states of the annealed fluid and then over the quenched
degrees of freedom of the matrix.  Using  the replica method one can  relate
the matrix averaged quantities to the thermodynamic quantities of the corresponding fully equilibrated model, called
a replicated model \cite{Given:92,Given_Stell:92,Given95,Rosinberg:94}. Significant  progress in the investigation of
partly-quenched systems
containing charges has been made  within the framework  of the replica Ornstein-Zernike  theory (see  review~\cite{Hribar_Lee:11}
and references cited therein). However,  the phase behaviour  has not been considered in these studies.

Recently, we have started a systematic investigation of the phase behaviour of ionic fluids confined in
a disordered porous medium, addressing the problem of the effects  of confinement on the vapour-liquid
phase diagrams \cite{HolPatPat16,HolPatPat17,HolPatPat17-2,PatPatHol17-3}. To this end, we have developed two  theoretical approaches: the  approach exploiting the concept of
ion association  \cite{HolKalyuzh91,Hol05} and
the other one which is based on the collective variable (CV) method
\cite{Yukhnovskii_Holovko,Cai-Pat-Mryg:05,Cai-Pat-Mryg,Patsahan_Mryglod_Caillol}. Both  approaches use the concept of a reference system.
In both cases, the
reference system is presented as  a two-component hard-sphere fluid  distributed in a  matrix of hard-sphere obstacles.
A pure analytical description of the thermodynamics of such a  reference system can be obtained from  a recent extension of
the scaled particle
theory (SPT) \cite{HolDong,PatHol11,HolPat12,HolPat13,HolPat15,HolovkoDong16,HolovkoPatsahanDong17}.  Within the framework
of the SPT approach,  a porous medium  is characterized by two types of  porosity. The first one is the so-called  geometrical
porosity $\phi_{0}$ characterizing   the free volume which is not  occupied  by matrix particles. The second porosity is
defined by the  chemical potential of a fluid  in the limit of infinite dilution and it is called a probe-particle porosity $\phi$.
This porosity characterises  the adsorption of a fluid particle  in an empty matrix. In the considered case of a two-component
hard-sphere fluid confined in a hard-sphere matrix, we have the  probe-particle porosity $\phi_{i}$ for each species $i$.
In order to describe correctly a fluid in the limit of  high density,
the new type of
porosity, $\phi^{*}$, has recently been introduced  \cite{HolPat12,HolovkoPatsahanDong17} which   corresponds to the maximum value of
fluid packing fraction  in a matrix. In the present paper, the reference system will be described using the SPT2b approximation
which  only includes porosities $\phi_{0}$ and $\phi_{i}$.

For a symmetric ionic fluid,   ion association can be treated  within the framework of either the associative mean spherical approximation
(AMSA) \cite{HolKalyuzh91,Hol05} or  the binding mean spherical approximation (BMSA) \cite{Blum95,Bernard96} which, in fact, are
identical. In \cite{Jiang02}, the AMSA theory  simplified in the spirit of a simple interpolation scheme introduced
by Stell and Zhou
\cite{Stell89}  was successfully  applied to the description of the vapour-liquid phase diagram of a symmetric ionic fluid in the bulk.
The approach was recently used for a symmetric ionic fluid in a disordered porous matrix \cite{HolPatPat17}.
In \cite{HolPatPat17-2}, using the  approach which
combines the SPT \cite{HolovkoDong16} and the
AMSA \cite{HolKalyuzh91,Hol05}  based on the so-called simplified mean spherical
approximation \cite{QinPrausnitz04}
we consider
a monovalent size-asymmetric PM  confined in a matrix of   hard-sphere or  overlapping hard sphere particles. However, the charge
asymmetry cannot be taken into account within this approach.

The  CV based approach   allows us to formulated the perturbation theory
using the extension of the SPT
for the description of  a reference system. The advantage of this approach is that one can derive an analytical expression
for the relevant chemical potential which includes the effects of correlations between ions
up to the third order and  takes into account both the charge- and size-asymmetry at the same level of approximation.
In \cite{PatPatHol17-3}, using this expression we calculate   the phase diagrams of a monovalent size-asymmetric PM with $\sigma_{+}/\sigma_{-}=1$,
$2$ and $3$ confined in hard-sphere matrices of different porosities and  different diameters of matrix obstacles.

Both the above-mentioned approaches produce qualitatively similar  dependencies of the vapour-liquid phase diagram  on the microscopic characteristics of
the matrix-ionic fluid model. It is worth noting that the results obtained
for the critical parameters of the PM demonstrate that the associative approach provides the better quantitative agreement with
simulation data in the bulk case. At the same time, these results  strongly depend on the definition of the association constant.

The present contribution is a continuation of our studies
described above. Here, we report the results  for  phase coexistence and critical parameters of a confined $2$:$1$
charge- and size-asymmetric PM obtained within the framework of the CV based theory. We analyse the trends of the critical temperature
and the critical density depending  on
the microscopic characteristics of the matrix-ionic fluid system and compare them with the charge-symmetric case.

The remainder of this paper is organized as follows. In Sec.~2, we briefly describe the theoretical background and present  the formulas
that are needed to calculate the phase diagrams.  The results are
presented and discussed in Sec.~3. We conclude in Sec.~4.

\section{Theoretical background}

Let us consider the general case of a two-component PM consisting of
$N_{+}$ hard spheres  of diameter
$\sigma_{+}$  carrying a charge $q_{+}=Zq$ and $N_{-}$ hard spheres of
diameter $\sigma_{-}$ carrying a charge $q_{-}=-q$. The ions are immersed in a structureless dielectric medium  with the dielectric constant $\varepsilon$.  The system is electrically neutral:
$\sum_{i=+,-}q_{i}\rho_{i}=0$ where
$\rho_{i}=N_{i}/V$ is the number density of the $i$th
species and $V$ is the system volume. The PM is characterized by the parameters of size and charge asymmetry:
\begin{equation}
\lambda=\frac{\sigma_{+}}{\sigma_{-}}\,, \qquad
Z=\frac{q_{+}}{|q_{-}|}\,.
\label{model_par}
\end{equation}

The ionic model is confined in a disordered porous matrix formed by hard spheres of diameter $\sigma_{0}$.
Then, the matrix-ionic fluid system is characterised by the ion-ion, ion-matrix and matrix-matrix interaction potentials.  The interaction
potentials between two ions are as follows:
\begin{eqnarray}
u_{ij}(r) = \left\{
                     \begin{array}{ll}
                     \infty, & r<\sigma_{ij}\\
                     \displaystyle\frac{q_{i}q_{j}}{\varepsilon r},&
                     r\geq \sigma_{ij}
                     \end{array}
              \right. \,,
            \qquad   i,j=+,-,
\label{int-PM}
\end{eqnarray}
where $\sigma_{ij}=\frac{1}{2}\left(\sigma_{i}+\sigma_{j}\right)$. The interaction potentials between an ion and a matrix particle and between two
matrix particles are described by the hard-sphere potentials:
\begin{eqnarray}
u_{0i}(r) = \left\{
                     \begin{array}{ll}
                     \infty, & r<\sigma_{0i}\\
                      0,& r\geq \sigma_{0i}
                     \end{array}
              \right. \,,           \qquad
 u_{00}(r) = \left\{
                     \begin{array}{ll}
                     \infty, & r<\sigma_{0}\\
                      0,& r\geq \sigma_{0}
                     \end{array}
              \right. \,.
\label{ion_matrix}
\end{eqnarray}
In (\ref{ion_matrix}), $\sigma_{0i}=\frac{1}{2}\left(\sigma_{0}+\sigma_{i}\right)$ and, in general,
$\sigma_{+}\neq\sigma_{-}\neq\sigma_{0}$. We introduce the parameter $\lambda_{0}$ which describes a size asymmetry between ions
and matrix particles defined as a size ratio of matrix obstacles and negatively charged ions:
\begin{equation}
\lambda_{0}=\sigma_{0}/\sigma_{-}.
\label{lambda0}
\end{equation}

Considering the matrix-ionic fluid system as a partly-quenched model we  can present the grand canonical potential of this system in the
following form \cite{HolPatPat16}:
\begin{equation}
-\beta\overline{\Omega}_{1}=\frac{1}{\Xi_{0}}\sum_{N_{0}\geq 0}\frac{z_{0}^{N_{0}}}{N_{0}!}
\int {\rm d}{\mathbf 0}
\exp(-\beta H_{00})\ln\Xi_{1},
\label{2.1}
\end{equation}
where
\begin{equation}
\Xi_{0} =\sum_{N_{0}\geq 0}\frac{z_{0}^{N_{0}}}{N_{0}!}\int {\rm d}{\mathbf 0}
\exp(-\beta H_{00})
\label{Xi0}
\end{equation}
is the grand  partition function for the prequenched medium
and
\begin{eqnarray}
\Xi_{1}&=&\sum_{N_{+}\ge 0}\sum_{N_{-}\ge 0}\frac{z_{+}^{N_{+}}}{N_{+}!}\frac{z_{-}^{N_{-}}}{N_{-}!}
\int {\rm d}{\mathbf 1^{+}}{\rm d}{\mathbf 1^{-}}\exp[-\beta(H_{0+}+H_{0-}+H_{++}
\nonumber \\
&&
+H_{+-}+H_{--})]
\label{Xi1}
\end{eqnarray}
is the matrix-dependent grand partition function. In (\ref{2.1})-(\ref{Xi1}), $z_{0}$, $z_{+}$, and $z_{-}$  are the activities
of the corresponding species. We write $H_{ij}$ for the sum of all pairwise interactions
between particles of species $i$ and species $j$ ($i,j=0,+,-$),  $\beta=1/k_{\rm{B}}T$, $k_{B}$ is the Boltzmann constant, $T$
is the absolute temperature.  Also,
${\rm d}{\mathbf 0}$,
${\rm d}{\mathbf 1^{+}}$, ${\rm d}{\mathbf 1^{-}}$  denote integration over all the positions of matrix particles and ions,
respectively.

Equation~(\ref{2.1}) can be simplified by using a replica method which consists in replacing the logarithm with an exponential. Thus,
we can rewrite (\ref{2.1}) as \cite{Given95}
\begin{equation}
-\beta\overline{\Omega}_{1}=\beta V\overline{P}=\lim_{s \to 0}\frac{{\rm d}}{{\rm d}s}\ln \Xi^{\rm{\rm{rep}}}(s),
\label{2.2}
\end{equation}
where
\begin{eqnarray}
\Xi^{\rm{rep}}(s)&=&\sum_{N_{0}\ge 0}\sum_{N_{1}^{+}\ge 0}\ldots
\sum_{N_{s}^{+}\ge 0}\sum_{N_{1}^{-}\ge 0}\ldots
\sum_{N_{s}^{-}\ge 0}\frac{z_{0}^{N_{0}}}{N_{0}!}\prod_{\alpha=1}^{s}\frac{z_{+,\alpha}^{N_{\alpha}^{+}}
z_{-,\alpha}^{N_{\alpha}^{-}}}{N_{\alpha}^{+}!N_{\alpha}^{-}!}
\nonumber \\
&&
\times\int {\rm d}{\mathbf 0} \prod_{\alpha=1}^{s}
{\rm d}{\mathbf 1}^{+}_{\alpha}
{\rm d}{\mathbf 1}^{-}_{\alpha}
\exp\{-\beta H_{00}-\beta\sum_{\alpha=1}^{s}[H_{0+}^{\alpha}+H_{0-}^{\alpha}]
\nonumber \\
&&
-\beta\sum_{\alpha,\beta=1}^{s}[H_{++}^{\alpha\beta}
+H_{+-}^{\alpha\beta}
+H_{--}^{\alpha\beta}]\}
\label{Ksi_rep}
\end{eqnarray}
is the grand partition function of a fully equilibrated  ($2s+1$)-component mixture, consisting
of the matrix  and of $s$ identical copies or replicas  of the two-component ionic fluid. Each pair of particles has the same
pairwise interaction in this replicated system as in the partly quenched  model except that a pair of ions from different
replicas has no interaction. Thus, the interaction potentials between matrix particles, matrix/fluid particles
and fluid/fluid particles read as
\begin{equation}
 u_{00}(r_{0}-r_{0}'), \quad u_{0A}^{\alpha}(r_{0}-r_{\alpha}^{A}), \quad
 u_{AB}^{\alpha\beta}(r_{\alpha}^{A}-r_{\beta}^{B})\delta_{\alpha\beta}.
 \label{int_poten}
\end{equation}
In the above equations, Latin indices denote fluid (ion)  species ($A,B=+,-$) and Greek indices denote
replicas ($\alpha,\beta=1,2,\ldots,s$).
The ($2s+1$)-component system with the interaction potentials (\ref{int_poten}) can be treated by using standard liquid state theories.

In \cite{PatPatHol17-3},  using  the CV method  we derive a functional representation of the grand partition function (\ref{Ksi_rep}).
For the model given by (\ref{int-PM})-(\ref{ion_matrix})  and (\ref{int_poten}), it can be presented as follows:
 \begin{eqnarray}
\Xi^{\rm{rep}}(s)=\Xi^{\rm{r}}[\bar{\nu}_{0},\bar{\nu}_{A}^{\alpha}]\int ({\rm d}\rho)({\rm d}\omega)
\exp\left[-\frac{\beta}{2}\sum_{{\mathbf k}}\widehat{U}(k)\widehat{\rho}_{{\mathbf k}}\widehat{\rho}_{-{\mathbf k}}\right.
\nonumber\\
\left.
+{\rm i}\sum_{{\mathbf k}}\widehat{\omega}_{{\mathbf k}}\widehat{\rho}_{{\mathbf k}}
+\sum_{n\geq 2}\frac{(-{\rm i})^{n}}{n!}\sum_{{\mathbf{k}}_{1},\ldots,{\mathbf{k}}_{n}}\widehat{{\mathfrak{M}}}_{n}\widehat{\omega}_{\mathbf k_{1}}\widehat{\omega}_{\mathbf k_{2}}\ldots\widehat{\omega}_{\mathbf k_{n}}\delta_{{\bf{k}}_{1}+\ldots +{\bf{k}}_{n}}
\right],
\label{Xi_matrix}
\end{eqnarray}
where  $\Xi^{r}$ is the grand partition function of a $(2s+1)$-component reference system with the renormalized partial chemical potentials
\begin{eqnarray*}
\bar{\nu}_{0}&=&\beta\mu_{0}-\ln\Lambda_{0}^{3},
\nonumber \\
\bar{\nu}_{A}^{\alpha}&=&\beta\mu_{A}^{\alpha}-\ln\Lambda_{A}^{3}+\frac{\beta}{2V}\sum_{{\mathbf
k}}\tilde u_{AA}^{\alpha\alpha(p)}(k),
\end{eqnarray*}
 $\mu_{0}$ and $\mu_{A}^{\alpha}$
are the chemical potentials of the corresponding species, $\Lambda_{0}$ and
$\Lambda_{A}$ are the  de Broglie thermal wavelengths,  $\tilde{u}_{AA}^{\alpha\alpha(p)}(k)$ is the
Fourier transform of the perturbative part of interaction potential ${u}_{AA}^{\alpha\alpha(p)}(r)$
\begin{equation}
{u}_{AA}^{\alpha\alpha(p)}(r)={u}_{AA}^{\alpha\alpha}(r)-{u}_{AA}^{\alpha\alpha(r)}(r),
\label{split}
\end{equation}
${u}_{AA}^{\alpha\alpha(r)}(r)$ is the interaction potential in the reference system.

$({\rm d}\rho)=({\rm
d}\rho_{0})({\rm d}\rho_{A}^{\alpha})$  ($({\rm d}\omega)=({\rm d}\omega_{0})({\rm d}\omega_{A}^{\alpha})$) denotes  volume elements
of the phase space of CVs $\rho_{{\mathbf k},0}$
and $\rho_{{\mathbf k},A}^{\alpha}$  ($\omega_{{\mathbf k},0}$ and $\omega_{{\mathbf k},A}^{\alpha}$). CVs $\rho_{{\mathbf k},0}$
and $\rho_{{\mathbf k},A}^{\alpha}$ describe the  fluctuation modes of the  number density of the matrix and fluid species, respectively ($\omega_{{\mathbf k},0}$ and $\omega_{{\mathbf k},A}^{\alpha}$ are
conjugate to $\rho_{{\mathbf k},0}$ and $\rho_{{\mathbf k},A}^{\alpha}$).

$\widehat{U}(k)$ is a symmetric $(2s+1)\times(2s+1)$
 matrix of elements:
\begin{eqnarray*}
u_{11}& =&0, \nonumber \\
u_{1i}&=&u_{i1} =0, \qquad
i\in E, \nonumber \\
u_{1i}&=&u_{i1} =0, \qquad
i\in O, \nonumber \\
u_{ii}&=&\widetilde{u}_{++}^{\alpha\alpha(p)}(k)=\widetilde{\varphi}_{++}(k), \qquad
i\in E,  \nonumber \\
u_{ii}&=&\widetilde{u}_{--}^{\alpha\alpha(p)}(k)=\widetilde{\varphi}_{--}(k), \qquad
i\in O,  \nonumber \\
u_{ij}&=&u_{ji}=\widetilde{u}_{+-}^{\alpha\alpha(p)}(k)=\widetilde{\varphi}_{+-}(k), \quad
i\in E, \quad  j=i+1, \nonumber \\
u_{ij}&=&0, \qquad  i\neq j, \quad   j\neq i+1,
\end{eqnarray*}
where the quantities with a ``tilde'' are Fourier transforms of perturbative parts of the corresponding interaction potentials
[see (\ref{split})]
and $E$ ($O$) are even (odd) numbers.
$\widehat{\rho}_{{\mathbf k}}$  indicates a column vector of elements $\rho_{{\mathbf k},0}$, $\rho_{{\mathbf k},+}^{1}$,
$\ldots$, $\rho_{{\mathbf k},+}^{s}$, $\rho_{{\mathbf k},-}^{1}$,
$\ldots$, $\rho_{{\mathbf k},-}^{s}$ and $\widehat{\omega}_{\mathbf k}$ is a row vector of elements $\omega_{{\mathbf k},0}$,
$\omega_{{\mathbf k},+}^{1}$,
$\ldots$, $\omega_{{\mathbf k},+}^{s}$, $\omega_{{\mathbf k},-}^{1}$,
$\ldots$, $\omega_{{\mathbf k},-}^{s}$.

$\widehat{{\mathfrak{M}}}_{n}$ is a  symmetric
$\underbrace{(2s+1)\times(2s+1)\times\ldots\times(2s+1)}_{n}$  matrix   whose elements are cumulants: the $n$th cumulant
coincides with the Fourier transform of the $n$-particle  truncated correlation function \cite{stell} of a reference
system. The  elements of  matrix $\widehat{{\mathfrak{M}}}_{2}$ are as follows:
\begin{eqnarray*}
{\mathfrak{M}}_{11}& =&{\mathfrak{M}}_{00}(k), \nonumber \\
{\mathfrak{M}}_{1i}&=&{\mathfrak{M}}_{i1} ={\mathfrak{M}}_{0+}(k), \quad
i\in E,
\qquad
{\mathfrak{M}}_{1i}={\mathfrak{M}}_{i1} ={\mathfrak{M}}_{0-}(k), \quad
 i\in O, \nonumber \\
{\mathfrak{M}}_{ii}&=&{\mathfrak{M}}_{++}^{11}(k), \quad
i\in E,
\qquad
{\mathfrak{M}}_{ii}={\mathfrak{M}}_{--}^{11}(k), \quad
i\in O,  \nonumber \\
{\mathfrak{M}}_{ij}&=&{\mathfrak{M}}_{ji}={\mathfrak{M}}_{+-}^{11}(k), \quad
i\in E, \quad  j\in O,  \quad j=i+1, \nonumber \\
{\mathfrak{M}}_{ij}&=&{\mathfrak{M}}_{ji}={\mathfrak{M}}_{++}^{12}(k), \quad
i,j\in E, \quad i\neq j, \nonumber \\
{\mathfrak{M}}_{ij}&=&{\mathfrak{M}}_{ji}={\mathfrak{M}}_{--}^{12}(k), \quad
i,j\in O, \quad i\neq j, \nonumber \\
{\mathfrak{M}}_{ij}&=&{\mathfrak{M}}_{ji}={\mathfrak{M}}_{+-}^{12}(k), \quad
i\in E, \quad  j\in O, \quad j\neq i+1,
\end{eqnarray*}
where
\begin{eqnarray*}
{\mathfrak{M}}_{00}(k)&=&\overline{\rho_{0}}\delta_{\mathbf{k}}+\overline{\rho_{0}}^{2}\tilde{h}_{00}^{(r)}(k), \quad
{\mathfrak{M}}_{0A}(k)=\overline{\rho_{0}}\,\overline{\rho_{A}}\tilde{h}_{0A}^{(r)}(k),\nonumber \\
{\mathfrak{M}}_{AB}^{\alpha\beta}(k)&=&\overline{\rho^{\alpha}_{A}}\delta_{AB}\delta_{\alpha\beta}\delta_{\mathbf{k}}+
\overline{\rho^{\alpha}_{A}}\,
\overline{\rho^{\beta}_{B}}
\tilde{h}_{AB}^{\alpha\beta(r)}(k),
\end{eqnarray*}
$\overline{\rho_{0}}=\langle N_{0}\rangle_{r}/V$, $\overline{\rho^{\alpha}_{A}}=\langle N_{A}^{\alpha}\rangle_{r}/V$, $\langle\ldots\rangle_{r}$
indicates the  average taken over the reference system and we put $\overline{\rho_{A}^{1}}=\overline{\rho_{A}^{2}}=\ldots=
\overline{\rho_{A}^{s}}=\overline{\rho_{A}}$.
$\tilde{h}_{\ldots}^{\ldots(r)}(k)$ is the Fourier transform of the corresponding pair correlation function of a $(2s+1)$-component reference system, $h_{AB}^{11(r)}(r)$ describes the correlations between particles within the same replica, whereas $h_{AB}^{12(r)}(r)$
 describes correlations between the particles from different replicas.

Taking into account the second order cumulants in (\ref{Xi_matrix}), after integration, we obtain the grand partition function  of the replicated system  in the Gaussian approximation
\begin{eqnarray}
 \frac{1}{V}\ln\Xi_{\rm{G}}^{\rm{rep}}(s)=\frac{1}{V}\ln\Xi^{r}
-\frac{1}{2V}\sum_{{\mathbf{k}}}\ln\left[\det(\widehat{U}\widehat{{\mathfrak{M}}}_{2}+\underline{1})\right].
\label{Ksi_rep_G}
\end{eqnarray}
From (\ref{Ksi_rep_G}), using a replica trick (\ref{2.2}), one derives an expression for the grand  potential of a
partly-quenched system in the Gaussian approximation
\begin{eqnarray}
-\beta\overline{\Omega}^{G}&=&-\beta\overline{\Omega}^{r}
-\frac{1}{2}\sum_{{\mathbf{k}}}\ln\left[\det(\widehat{\Phi}_{2}\widehat{{\mathfrak{M}}}_{2}^{c}+\underline{1})\right]
-\frac{1}{2}\sum_{{\mathbf{k}}}\frac{1}{\det(\widehat{\Phi}_{2}\widehat{{\mathfrak{M}}}_{2}^{c}
+\underline{1})}
\nonumber \\
&&
\times\left\{\det(\widehat{\Phi}_{2})\left({\mathfrak{M}}_{++}^{c}{\mathfrak{M}}_{--}^{b}+
{\mathfrak{M}}_{--}^{c}{\mathfrak{M}}_{++}^{b}-2{\mathfrak{M}}_{+-}^{c}{\mathfrak{M}}_{+-}^{b}\right)\right.
\nonumber \\
&&
\left.
+\sum_{A,B=+.-}\beta\widetilde{\varphi}_{AB}(k){\mathfrak{M}}_{AB}^{b}
\right\},
\label{Omega_PQ_0}
\end{eqnarray}
 where $\overline{\Omega}^{r}$ is the grand  potential of the reference system consisting of  a two-component hard-sphere fluid confined
in a hard-sphere matrix.
Matrix  $ \widehat{\Phi}_{2}$ is of the form:
\[
 \widehat{\Phi}_{2}=
\begin{pmatrix}
  \beta\widetilde{\varphi}_{++}(k) & \beta\widetilde{\varphi}_{+-}(k) \\
  \beta\widetilde{\varphi}_{+-}(k) & \beta\widetilde{\varphi}_{--}(k)
\end{pmatrix}.
\]
In the case of the Weeks-Chandler-Andersen  regularization scheme \cite{wcha}, the Fourier transforms of the Coulomb interaction potentials
read
\begin{eqnarray*}
\beta\widetilde{\varphi}_{++}(k)&=&\frac{4\pi Z\sigma_{+-}^{3}}{T^{*}(1+\delta)}\frac{\sin(x(1+\delta))}{x^{3}},
\\
\beta\widetilde{\varphi}_{--}(k)&=&\frac{4\pi\sigma_{+-}^{3} }{T^{*}Z(1-\delta)}\frac{\sin(x(1-\delta))}{x^{3}},
\\
\beta\widetilde{\varphi}_{+-}(k)&=&-\frac{4\pi\sigma_{+-}^{3} }{T^{*}}\frac{\sin(x)}{x^{3}},
\end{eqnarray*}
where   $T^{*}=k_{B}T\sigma_{+-}/(q^{2}Z)$ is the dimensionless temperature,
$x=k\sigma_{+-}$,  $\sigma_{+-}=(\sigma_{+}+\sigma_{-})/2$, and
\begin{equation}
 \delta=\frac{\lambda-1}{\lambda+1}.
\label{delta}
\end{equation}

${\mathfrak{M}}_{AB}^{c}$ and ${\mathfrak{M}}_{AB}^{b}$ are elements of the matrices
\begin{equation}
 \widehat{{{\mathfrak{M}}}}_{2}^{c}=
\begin{pmatrix}
  {\mathfrak{M}}_{++}^{c}(k) & {\mathfrak{M}}_{+-}^{c}(k) \\
  {\mathfrak{M}}_{+-}^{c}(k) & {\mathfrak{M}}_{--}^{c}(k) \\
\end{pmatrix}, \qquad
\widehat{{{\mathfrak{M}}}}_{2}^{b}=
\begin{pmatrix}
  {\mathfrak{M}}_{++}^{b}(k) & {\mathfrak{M}}_{+-}^{b}(k) \\
  {\mathfrak{M}}_{+-}^{b}(k) &{\mathfrak{M}} _{--}^{b}(k) \\
\end{pmatrix}.
\label{M-matrix}
\end{equation}
In (\ref{M-matrix}), superscripts ``c'' and ``b''  denote  the connected and blocking parts of the cumulants
${\mathfrak{M}}_{AB}$ (structure factors of the reference system):
\begin{eqnarray}
{\mathfrak{M}} _{AB}(k)={\mathfrak{M}}_{AB}^{c}(k)+{\mathfrak{M}}_{AB}^{b}(k)
=\rho_{A}\delta_{AB} +\rho_{A}\rho_{B}\widetilde{h}_{AB}^{r}(k),
\label{M_AB_s0}
\end{eqnarray}
where
\begin{eqnarray*}
 {\mathfrak{M}}_{AB}^{c}(k)&=&\rho_{A}\delta_{AB}
 +\rho_{A}\rho_{B}\widetilde{h}_{AB}^{r,c}(k), \nonumber \\
 \widetilde{h}_{AB}^{r,c}(k)&=&\lim_{s \to 0}[\widetilde{h}_{AB}^{11(r)}(k)-\widetilde{h}_{AB}^{12(r)}(k)],
\end{eqnarray*}
and
\begin{eqnarray*}
{\mathfrak{M}}_{AB}^{b}(k)=\rho_{A}\rho_{B}\widetilde{h}_{AB}^{r,b}(k), \qquad
\widetilde{h}_{AB}^{r,b}(k)=\lim_{s \to 0}\widetilde{h}_{AB}^{12(r)}(k).
\end{eqnarray*}
In (\ref{M_AB_s0}),  $\widetilde{h}_{AB}^{r}(k)=\widetilde{h}_{AB}^{r,c}
+\widetilde{h}_{AB}^{r,b}$  is the Fourier transform of the partial pair correlation function with
$\widetilde{h}_{AB}^{r,c(b)}$ being its connected (blocking) part \cite{Given_Stell:92,Given_Stell:92_2}.

For $\overline{\mathfrak{M}}_{00}$ and $\overline{\mathfrak{M}}_{0A}$, we have
\begin{eqnarray*}
\overline{\mathfrak{M}}_{00}(k)&=&\rho_{0}+\rho_{0}^{2}\widetilde{h}_{00}^{r}(k),
 \qquad
 \widetilde{h}_{00}^{r}(k)=\lim_{s \to 0}\widetilde{h}_{00}^{(r)}(k) \nonumber \\
\overline{\mathfrak{M}}_{0A}(k)&=&\rho_{0}\rho_{A}\widetilde{h}_{0A}^{r}(k),
 \qquad
 \widetilde{h}_{0A}^{r}(k)=\lim_{s \to 0}\widetilde{h}_{0A}^{(r)}(k),
\end{eqnarray*}
where $\widetilde{h}_{00}^{r}(k)$ and $\widetilde{h}_{0A}^{r}(k)$ are Fourier transforms of the matrix-matrix and
matrix-fluid correlation functions in a partly-quenched reference system.

Based on (\ref{Omega_PQ_0}), we determine the chemical potential, or, more precisely, a linear
combination of the partial chemical potentials, conjugate to the order parameter
of the vapour-liquid critical point using the method  proposed for the bulk PM
\cite{patsahan-mryglod-patsahan:06,PatPat09,Patsahan_Patsahan:10}. As a result, we obtain
\begin{eqnarray}
\nu_{1}&=&\frac{\nu_{+}+Z\nu_{-}}{\sqrt{1+Z^{2}}}=\nu_{1}^{0}+\frac{\sqrt{1+Z^{2}}}{2\left[{\mathfrak{M}}_{++}^{c}+
2Z{\mathfrak{M}}_{+-}^{c}
+Z^{2}{\mathfrak{M}}_{--}^{c}\right]}\times
\nonumber
\\
&& \times
\frac{1}{V}\sum_{{\mathbf
k}}\frac{1}{{\rm
det}\,[\widehat{\Phi}_{2}\widehat{{\mathfrak{M}}}_{2}^{c}+\underline{1}]}
\left[\beta\tilde\varphi_{++}(k){\cal F}_{1}
+\beta\tilde\varphi_{--}(k){\cal F}_{2}+\right.
\nonumber \\
&&
\left.+2\beta\tilde\varphi_{+-}(k){\cal F}_{3}\right],
\label{nu1}
\end{eqnarray}
where
\begin{equation}
\nu_{1}^{0}=\frac{\nu_{+}^{r}+Z\nu_{-}^{r}}{\sqrt{1+Z^{2}}}-\frac{1}{2\sqrt{1+Z^{2}}}\frac{1}{V}
\sum_{{\mathbf k}}\left( \beta\tilde\phi_{++}(k)+Z\beta\tilde\phi_{--}(k)\right)
\label{nu1_0}
\end{equation}
is the mean-field part of $\nu_{1}$. In (\ref{nu1_0}), $\nu_{A}^{r}$ is the chemical potential
of the corresponding species in the reference system. Analytical expressions for $\nu_{+}^{r}$ and $\nu_{-}^{r}$ obtained in the so-called SPT2b approximation
are given in Appendix~A. It should be emphasized that, in addition to the second-order cumulants  ${\mathfrak{M}}_{AB}^{c}$,
Eq.~(\ref{nu1}) includes the connected parts of the third-order
cumulants ${\mathfrak{M}}_{ABC}^{c}$:
\begin{eqnarray}
{\cal F}_{1}&=&{\mathfrak{M}}_{+++}^{c}+Z{\mathfrak{M}}_{++-}^{c}, \quad {\cal
F}_{2}={\mathfrak{M}}_{+--}^{c}+Z{\mathfrak{M}}_{---}^{c},\nonumber \\
{\cal F}_{3}&=&{\mathfrak{M}}_{++-}^{c}+Z{\mathfrak{M}}_{+--}^{c}.
\label{F-3}
\end{eqnarray}
All cumulants in (\ref{nu1}) are approximated by their values in the long-wavelength limit. Then, analytical expressions for
${\mathfrak{M}}_{AB}^{c}$ and ${\mathfrak{M}}_{ABC}^{c}$ can be obtained from (\ref{nu_+_hs})-(\ref{k+3}) using the
Kirkwood-Buff equations \cite{kirkbuf}.   The  formulas linking   ${\mathfrak{M}}_{AB}^{c}$ and
${\mathfrak{M}}_{ABC}^{c}$  to derivatives of the chemical potentials are given in \cite{PatPatHol17-3}.
It is worth noting that equation (\ref{nu1}) is derived in the case $\nu_{2}=\nu_{2}^{0}$ where
$\nu_{2}=(Z\nu_{+}-\nu_{-})/\sqrt{1+Z^{2}}$.

In the following section, Eqs.~(\ref{nu1})-(\ref{F-3}) are used to study the vapour-liquid equilibria of the $2$:$1$ PM
confined in a disordered hard-sphere matrix.

\section{Results and discussion}
The CV theory presented in the previous section is applied to build phase diagrams of the vapour-liquid  phase
transition for a $2$:$1$ size-asymmetric  PM fluid ($Z=2$) in a disordered hard-sphere matrix of different geometrical porosities $\phi_{0}$ ($\phi_{0}=0.85-1.0$) and different sizes of matrix particles  ($\lambda_{0}=\sigma_{0}/\sigma_{-}=1.0-3.0)$.
The size-asymmetry ratio of fluid particles $\lambda=\sigma_{+}/\sigma_{-}$ is taken as $1/3$, $1/2$, $1$, $2$ and $3$. It means
that we describe the case of cations larger than anions and when they are smaller than anions.
The PM fluid of equally sized ions is also considered. For comparison, we  present the results obtained using the same formalism for a symmetrically charged PM fluid ($Z=1$).

As it  was mentioned in Introduction, we distinguish two types of porosity: the geometrical porosity $\phi_0$ and the probe-particle porosity $\phi_{i}$ for each species $i$ ($i=+,-$). A principal difference between them consists in the fact that the probe-particle porosities $\phi_{+}$ and $\phi_{-}$ take into account the size of adsorbate particles,
while the porosity $\phi_0$ is an adsorbate-independent characteristic. Both types of  porosities are important. The geometrical porosity $\phi_{0}$ defines a ``bare'' pore volume of the matrix and it can be considered as a more general characteristic. For the hard-sphere matrix, $\phi_{0}=1-\eta_{0}$, where $\eta_{0}=\pi\rho_{0}\sigma_{0}^{3}/6$, $\rho_{0}=N_{0}/V$ is the
number density of  matrix particles. The porosities $\phi_{+}$ and $\phi_{-}$  can be calculated using the equations in Appendix~A.

The phase diagrams are built by curves describing a coexistence between the vapour and liquid phases,
the density of which are obtained by the Maxwell construction for the chemical potential (\ref{nu1}).
Here, we  introduce the  reduced units for the temperature and for  the density which are conventionally used in
the works dealing with the phase behavior of an asymmetric PM in the bulk state
(see for example \cite{Romero-Enrique:00}), i.e.,
\begin{equation*}
T^{*}=\frac{k_{\mathrm{B}}T\sigma_{+-}}{q^{2}Z}, \qquad
\rho^{*}=\rho\sigma_{+-}^{3},
\end{equation*}
where $\rho=\rho_{+}+\rho_{-}$ is the total ionic number density.  For the bulk case,  the present approximation reproduces
a correct trend  of $T^{*}_{c}$ with size and charge asymmetry as compared with simulations: $T^{*}_{c}$ decreases when $\lambda$ and $Z$ increase \cite{Patsahan_Patsahan:10}. Regarding  $\rho^{*}_{c}$, its trend
with $\lambda$ at the fixed $Z$ qualitatively agrees  with simulations while it shows the opposite trend  with charge asymmetry
in the equisized case \cite{Patsahan_Patsahan:10}.
It follows from \cite{patsahan-mryglod-patsahan:06} that the taking into account of the correlation effects of the
higher order than the third order leads to the correct
trend of the critical density with charge asymmetry.

\begin{figure}[ht]
\begin{center}
\includegraphics[clip,width=0.47\textwidth,angle=0]{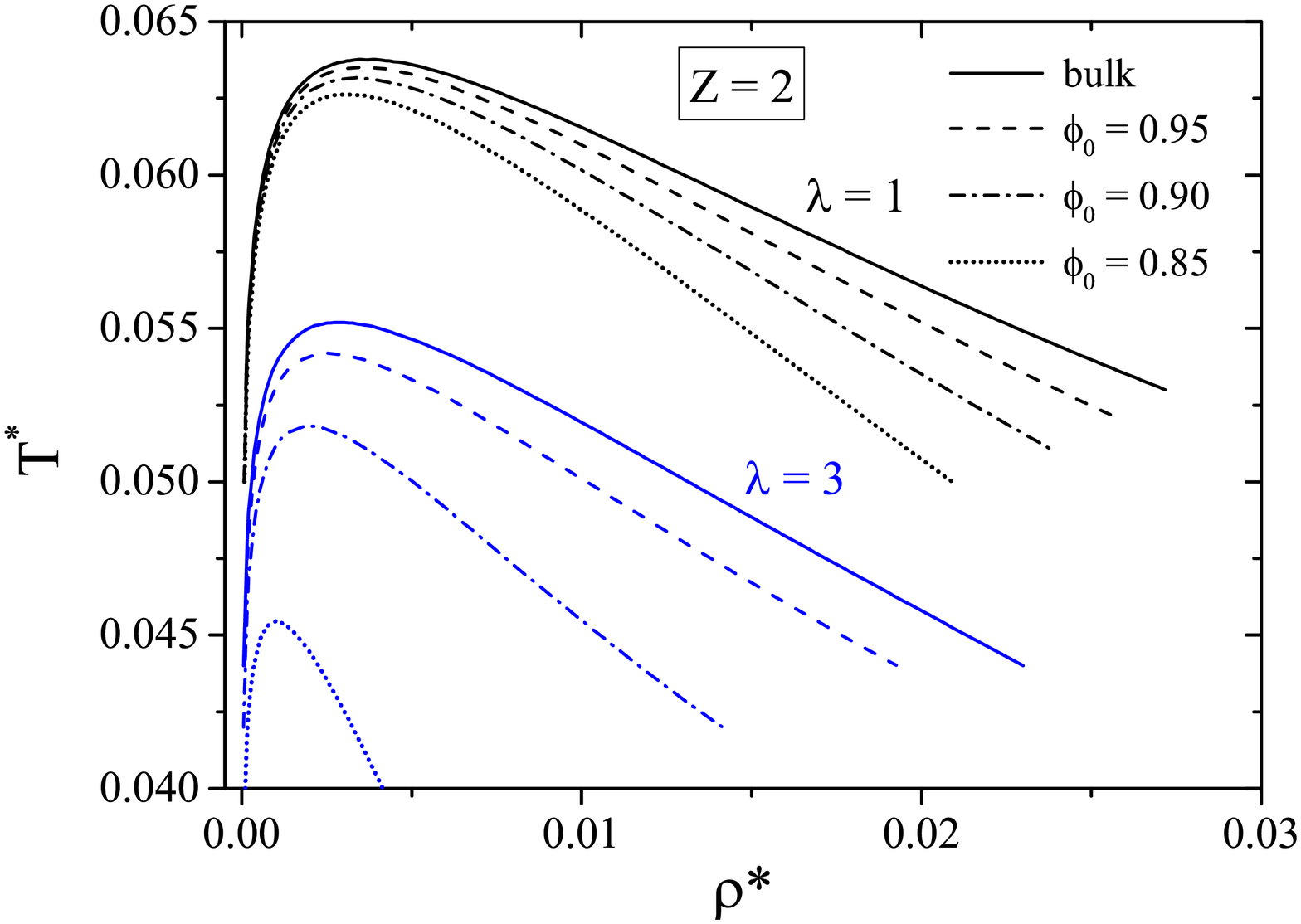}
\includegraphics[clip,width=0.47\textwidth,angle=0]{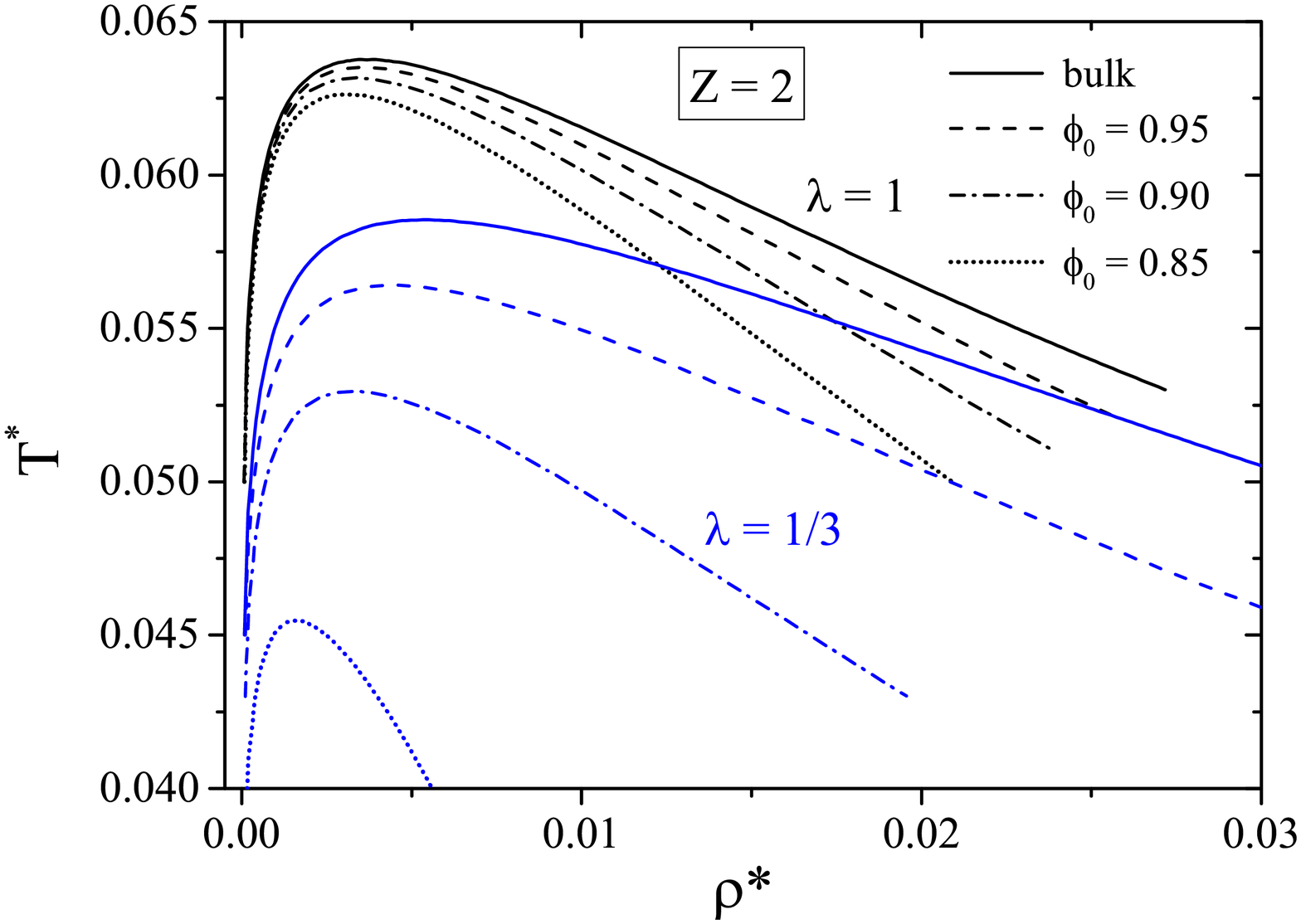} \caption{\label{fig:phase}
Vapour-liquid phase diagrams of a $2$:$1$ PM fluid confined in a matrix of different porosities, $\phi_{0}$,
but at the fixed size of matrix particles $\lambda_{0}=1$. Black lines denote an equisized PM fluid ($\lambda=1$)
and blue lines correspond to a size-asymmetric PM fluid ($\lambda=3$ -- left panel, $\lambda=1/3$ -- right panel).}
\end{center}
\end{figure}

We calculate a number of  phase diagrams  in the ($T^{*}\--\rho^{*}$) plane
for  different system parameters.
As an example, we present some of them in Fig.~\ref{fig:phase} to demonstrate that for $Z=2$ the general trend remains similar to the case of $Z=1$ studied
in our previous work \cite{PatPatHol17-3}, i.e., a decrease of the porosity $\phi_{0}$ leads to a shift of
the phase coexistence region toward lower temperatures and densities, and simultaneously this region gets narrower.
On the other hand, the phase coexistence region of a $2:1$ PM fluid can get broader due to a size-asymmetry,
as it is observed in Fig.~\ref{fig:phase}~(right panel) for $\lambda=1/3$ in the bulk and in the matrix of high porosity $\phi_{0}=0.95$.
However, a decrease of the matrix porosity suppresses this effect.

\begin{figure}[ht]
\begin{center}
\includegraphics[clip,width=0.47\textwidth,angle=0]{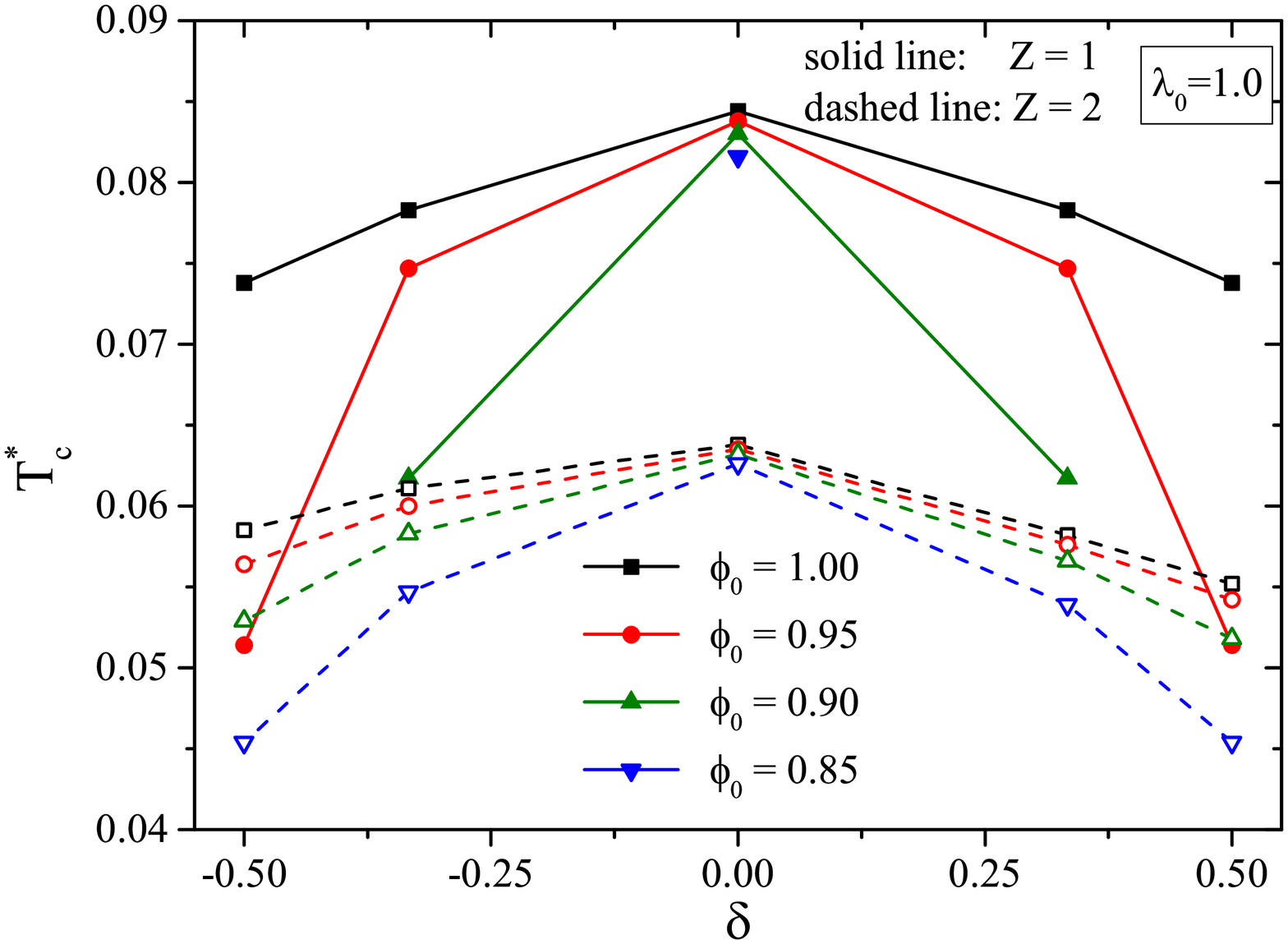}
\includegraphics[clip,width=0.47\textwidth,angle=0]{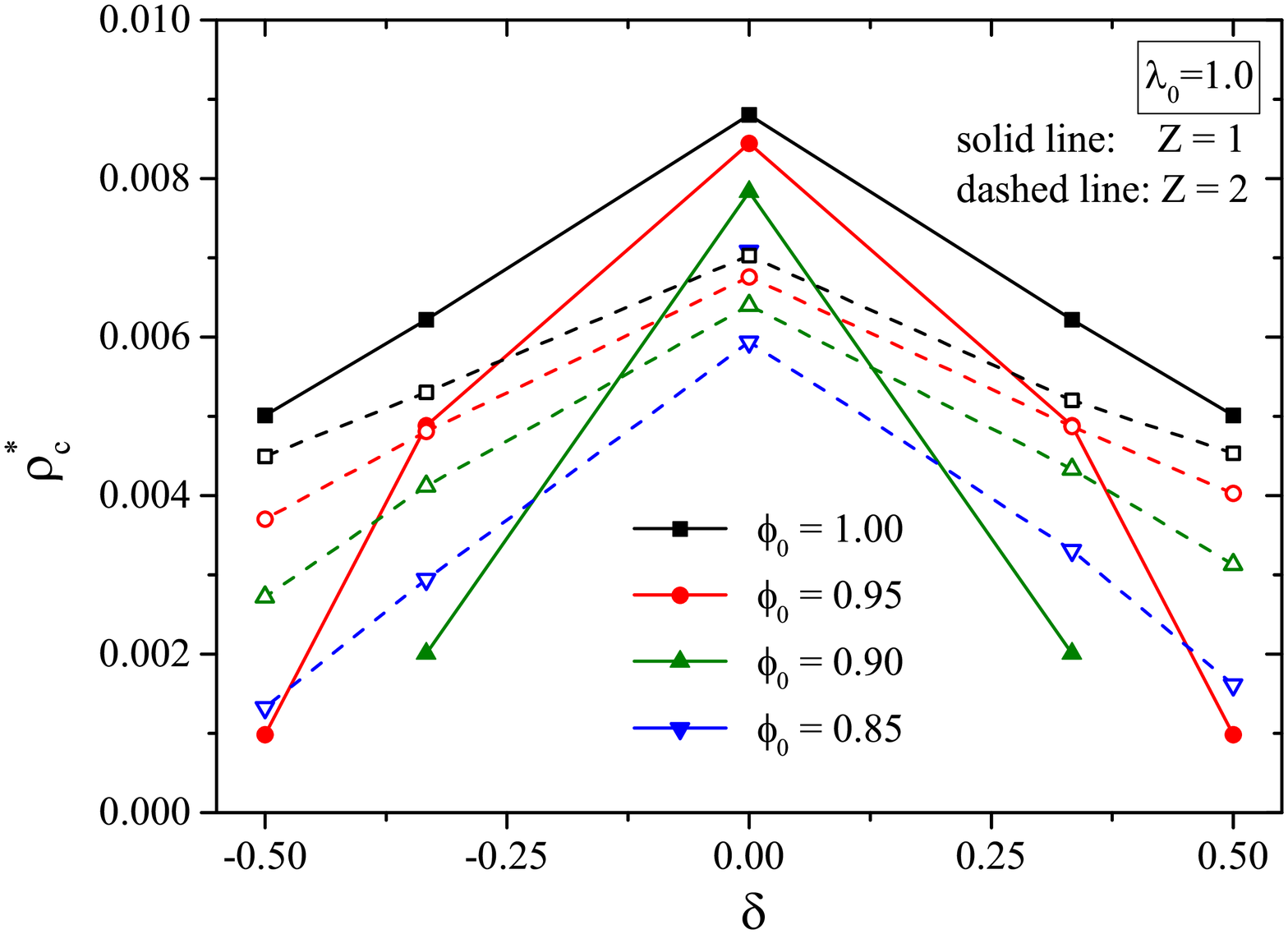} \caption{\label{fig:critDelta}
Critical temperature $T^{*}_{c}$ (left panel) and critical density $\rho^{*}_{c}$ (right panel) of a $Z$:$1$ PM fluid
as a function of size asymmetry ratio $\delta$ in the bulk ($\phi_{0}=1.0$)
and in a matrix of different porosities $\phi_{0}$, but at the fixed size of matrix particles
$\lambda_{0}=1$.
Solid lines correspond to the case  $Z=1$, dashed lines -- $Z=2$.}
\end{center}
\end{figure}

\begin{figure}[ht]
\begin{center}
\includegraphics[clip,width=0.47\textwidth,angle=0]{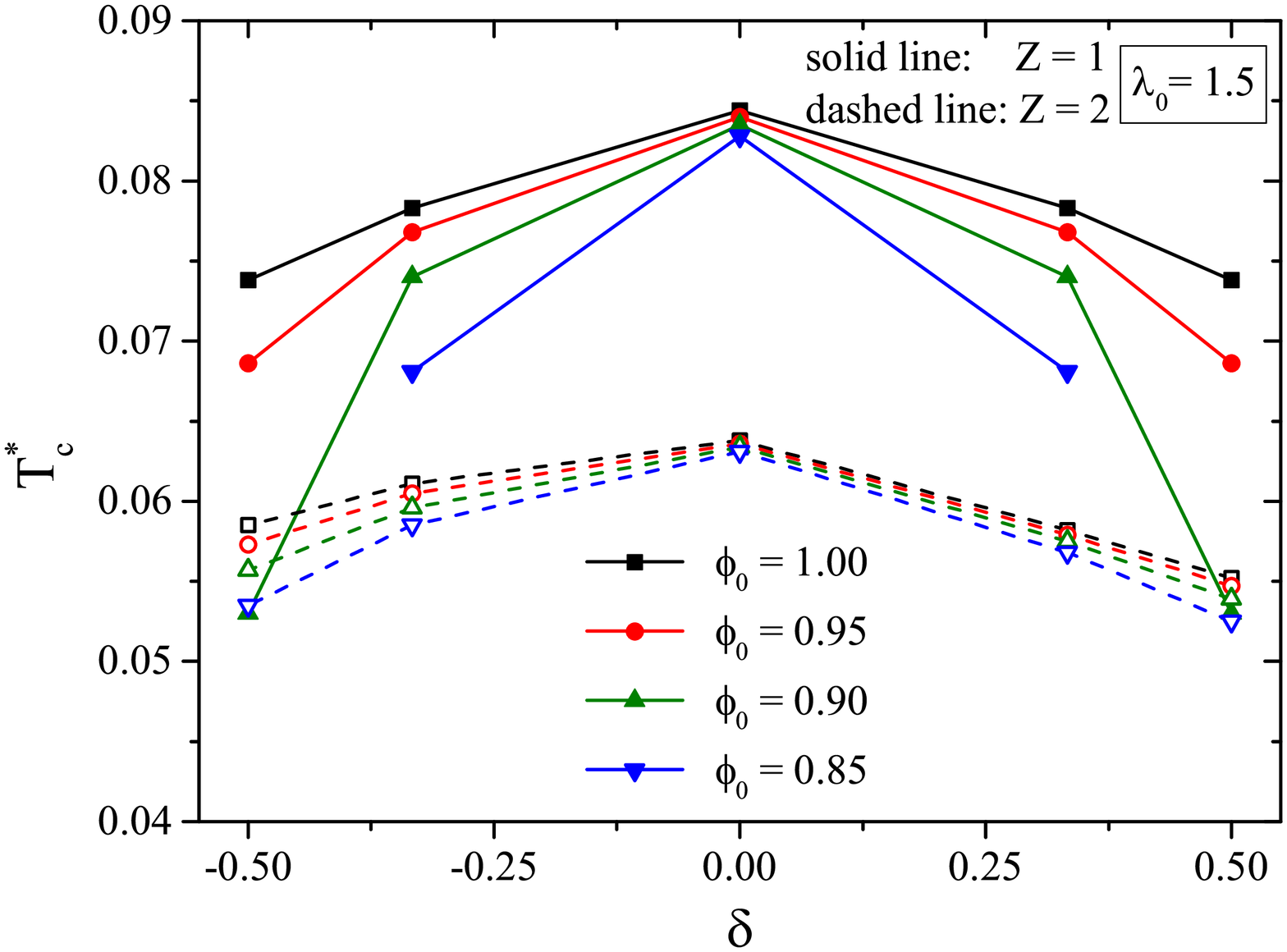}
\includegraphics[clip,width=0.47\textwidth,angle=0]{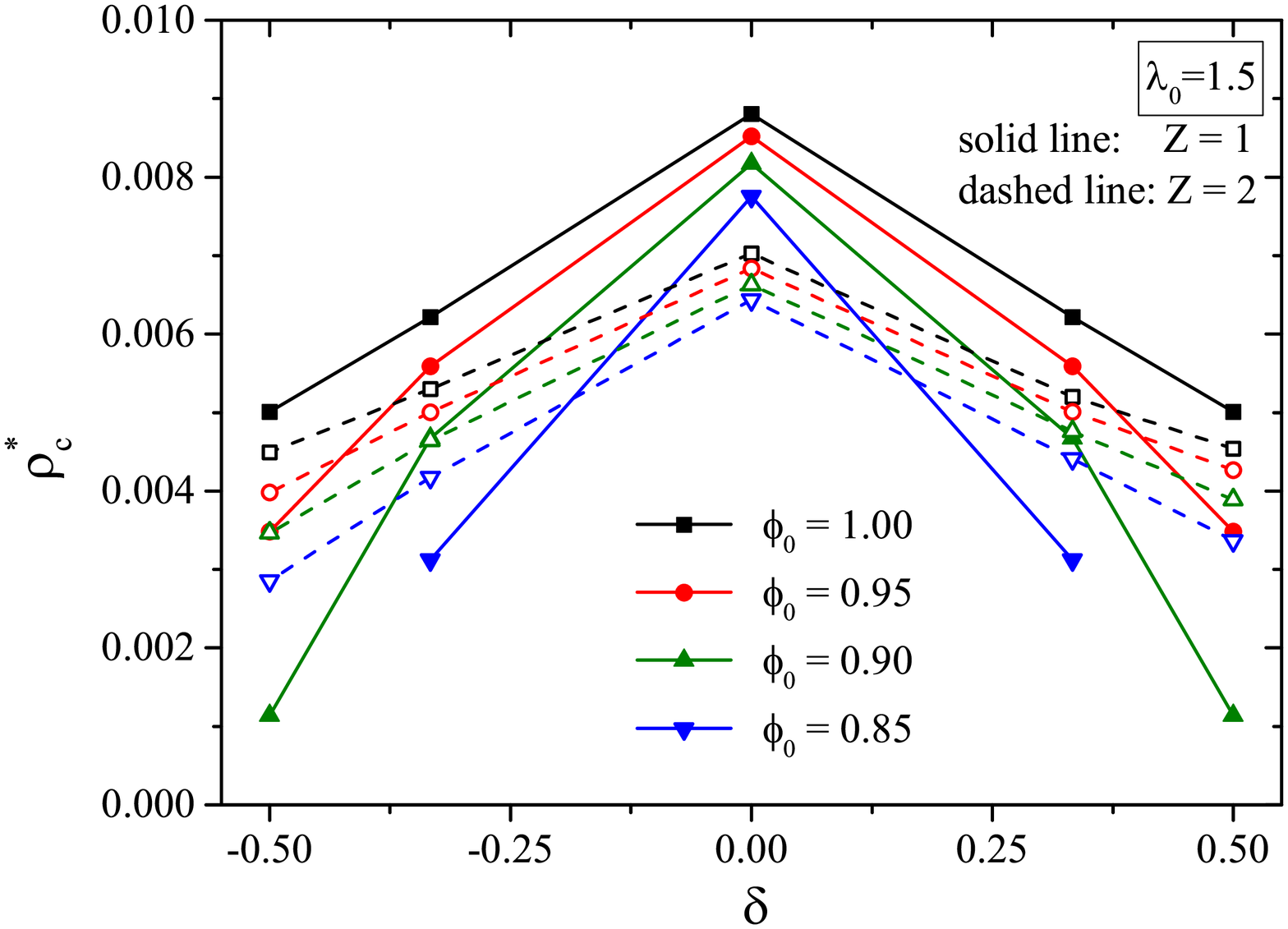} \caption{\label{fig:critDeltaL1.5}
The same as in Fig.~\ref{fig:critDelta}, but for $\lambda_{0}=1.5$.}
\end{center}
\end{figure}

Based on the phase diagrams,  we  analyse the effect of disordered matrix on the critical parameters $T^{*}_{c}$ and $\rho^{*}_{c}$
defining the point  for which the maxima and minima of the van der Waals loops coalesce.
First, we inspect how a matrix presence changes the  effects of ion size asymmetry.
For this purpose, we fix the size of matrix particles $\lambda_0$ and consider different values of the  size ratio  of positively and
negatively charged particles  confined in the matrix of different porosities $\phi_{0}$.
In Fig.~\ref{fig:critDelta} (left panel), one can observe the dependencies of $T^{*}_{c}$ and $\rho^{*}_{c}$ on the parameter of
ion size asymmetry $\delta$
given by (\ref{delta}). For $Z=1$, these dependencies
are characterized by the ideal symmetry of $T^{*}_{c}$ with respect to $\delta=0$.
However, such a behaviour is not seen for $Z=2$, and  it is especially noticeable in the bulk case ($\phi_{0}=1$).
On the other hand, a decrease of the porosity makes the dependencies for $Z=2$ more symmetric.
Similarly,  the dependency of $\rho^{*}_{c}$ on $\delta$ is symmetric with respect to $\delta=0$ for $Z=1$ and
 it is asymmetric for $Z=2$ (Fig.~\ref{fig:critDelta}, right panel).
Contrary to the critical temperature, an  asymmetric behaviour of $\rho^{*}_{c}$  with respect to $\delta=0$ for $Z=2$ becomes
more essential if the matrix porosity decreases.

\begin{figure}[ht]
\begin{center}
\includegraphics[clip,width=0.47\textwidth,angle=0]{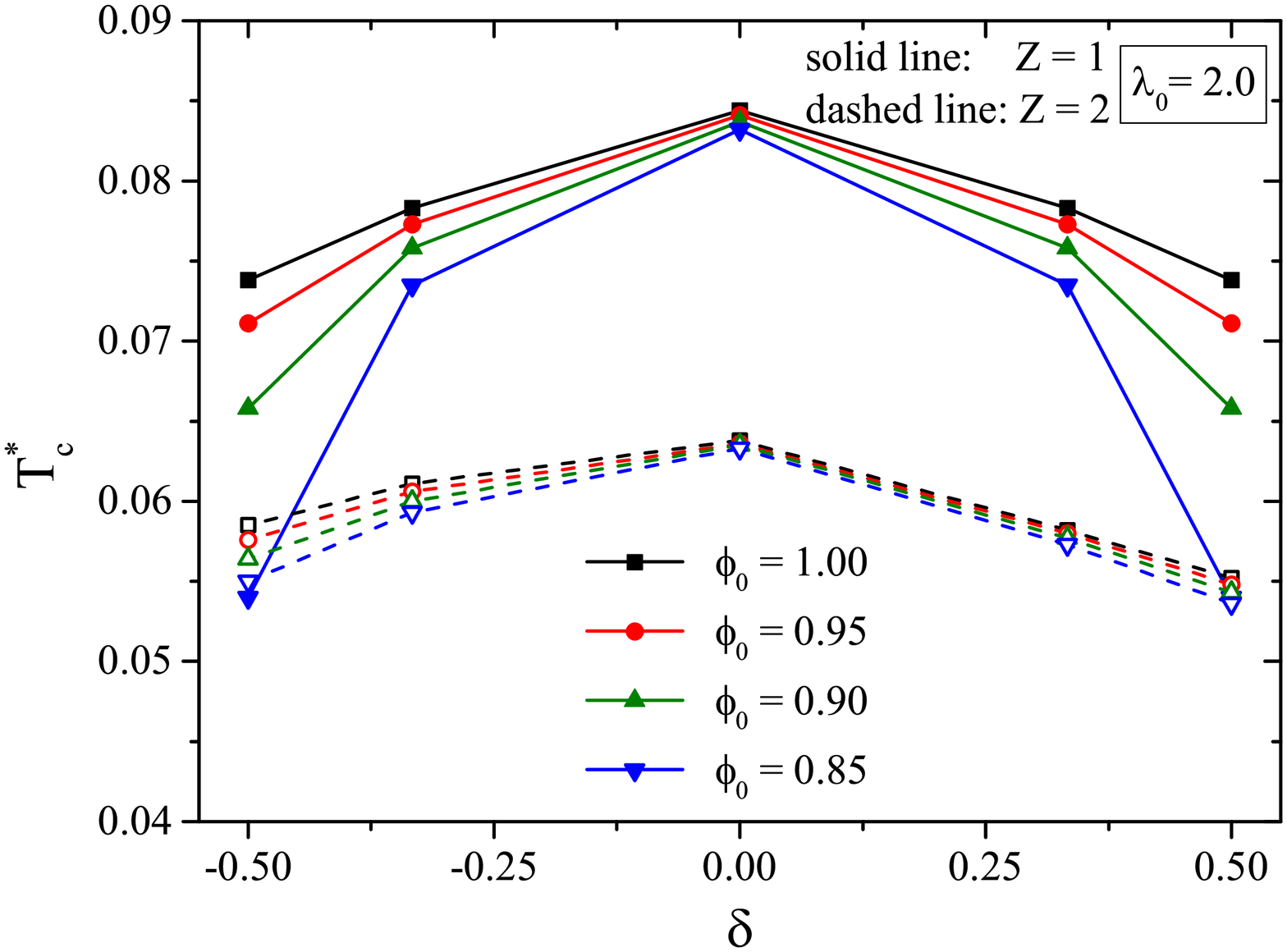}
\includegraphics[clip,width=0.47\textwidth,angle=0]{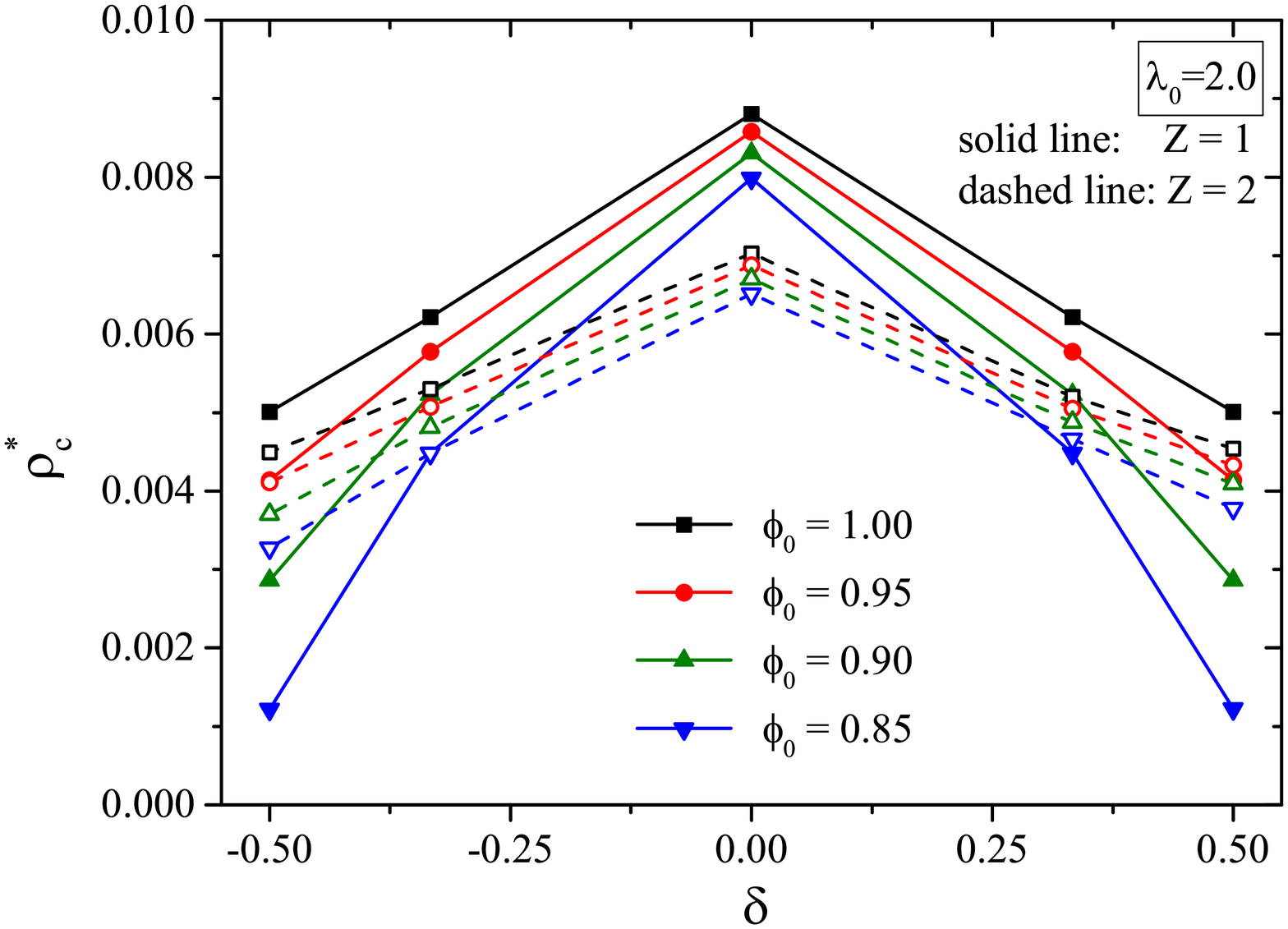} \caption{\label{fig:critDeltaL2.0}
The same as in Fig.~\ref{fig:critDelta}, but for $\lambda_{0}=2.0$.}
\end{center}
\end{figure}

\begin{figure}[ht]
\begin{center}
\includegraphics[clip,width=0.47\textwidth,angle=0]{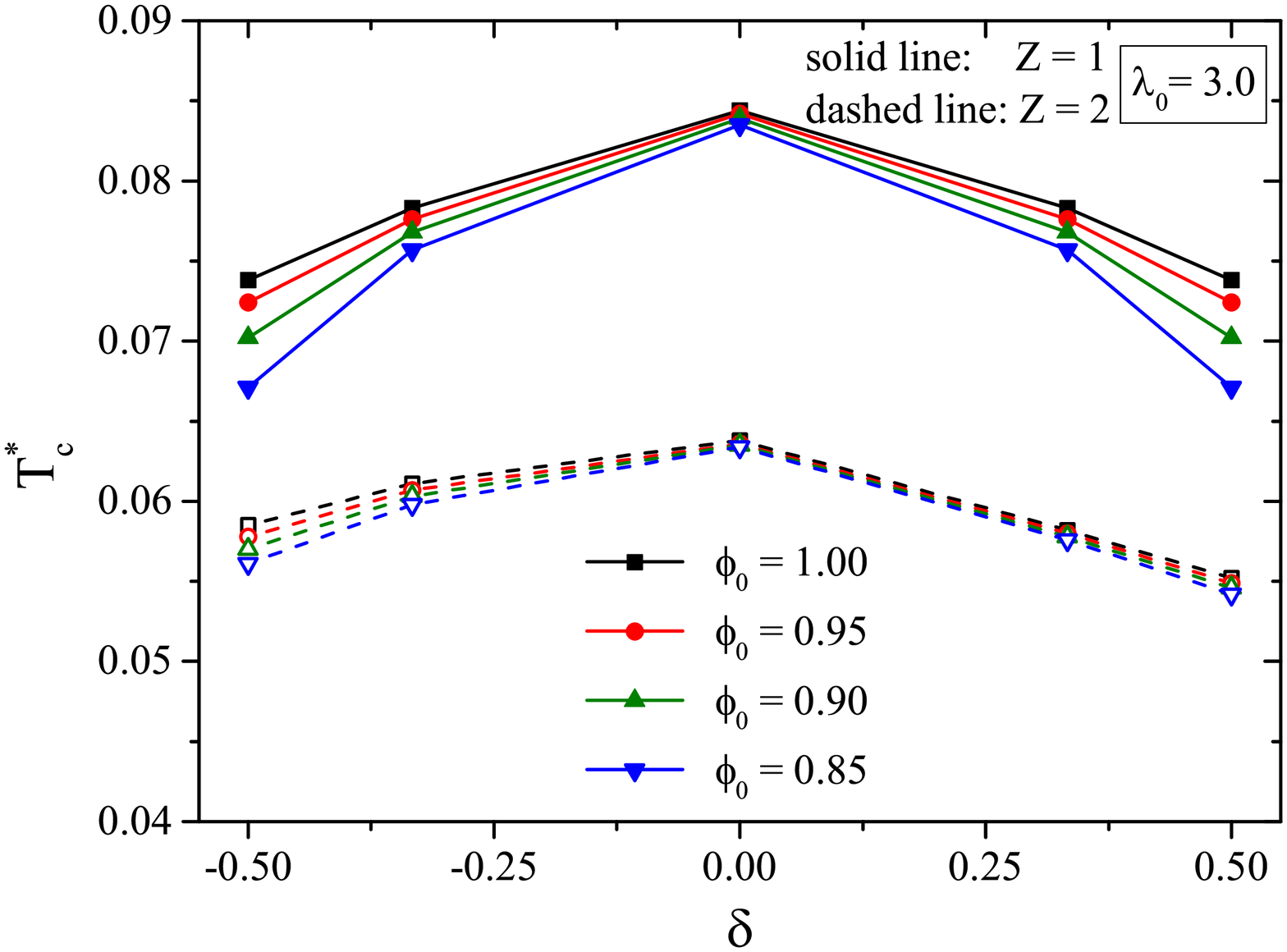}
\includegraphics[clip,width=0.47\textwidth,angle=0]{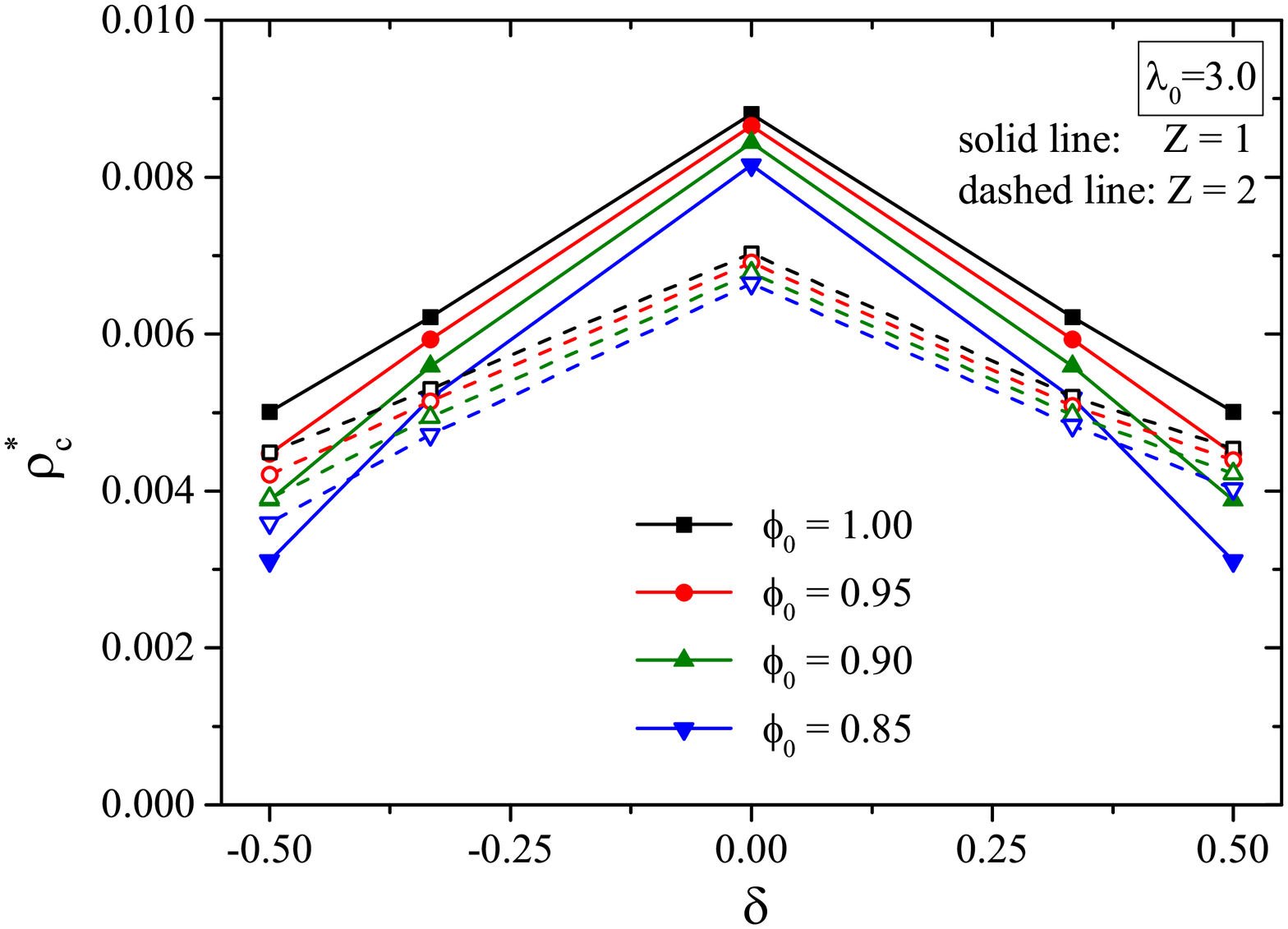} \caption{\label{fig:critDeltaL3.0}
The same as in Fig.~\ref{fig:critDelta}, but for $\lambda_{0}=3.0$.}
\end{center}
\end{figure}

Both the critical temperature and the critical density are affected by the matrix porosity.
 It is shown for $Z=1$  \cite{PatPatHol17-3} that the smaller is the porosity the lower is the critical temperature.
The same trend is observed for the critical density if $Z=1$. In general, a similar behaviour of $T^{*}_{c}$ and $\rho^{*}_{c}$ is found
in the case of $Z=2$  (see Figs.~\ref{fig:critDelta}--\ref{fig:critDeltaL3.0}). Nevertheless, there are some differences in
the behaviour of the critical parameters with size asymmetry for $Z=1$ and $Z=2$.
As one can see from Figs.~\ref{fig:critDelta}--\ref{fig:critDeltaL3.0},  both critical parameters of a charge-asymmetric PM fluid
are less  sensitive to the ion size asymmetry, hence they decrease slower with $\delta$ for $Z=2$ in comparison
with $Z=1$. Despite the matrix presence  strengthens a decrease of $T^{*}_{c}$ and $\rho^{*}_{c}$ with $\lambda$,
this effect is more pronounced in the case of $Z=1$.
On the other hand, an increase in the size of matrix particles at the fixed porosity weakens the effects of  matrix presence.
Figs.~\ref{fig:critDeltaL1.5}--\ref{fig:critDeltaL3.0}  show that the critical temperatures of a PM fluid confined in matrices
of large particles are getting closer to the corresponding values obtained in the bulk.
It is especially noticeable for $T^{*}_{c}$ when $Z=2$. Unlike a charge-symmetric PM,  a vapour-liquid
phase coexistence in a charge-asymmetric fluid is found for all matrix characteristics ($\lambda_{0}$ and $\phi_{0}$)
considered in this study (see Figs.~\ref{fig:critDelta}--\ref{fig:critDeltaL1.5}).

\begin{figure}[ht]
\begin{center}
\includegraphics[clip,width=0.47\textwidth,angle=0]{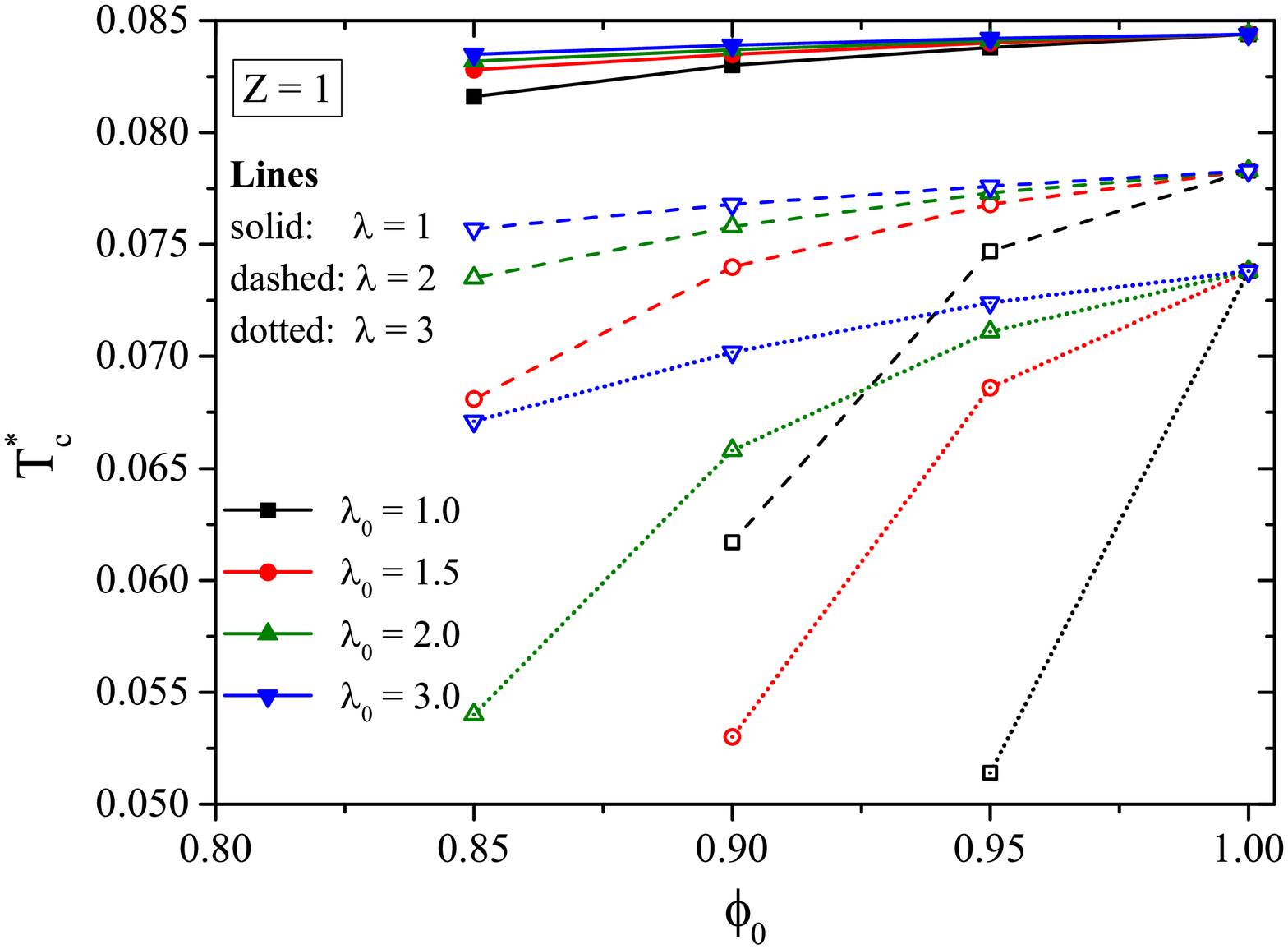}
\includegraphics[clip,width=0.47\textwidth,angle=0]{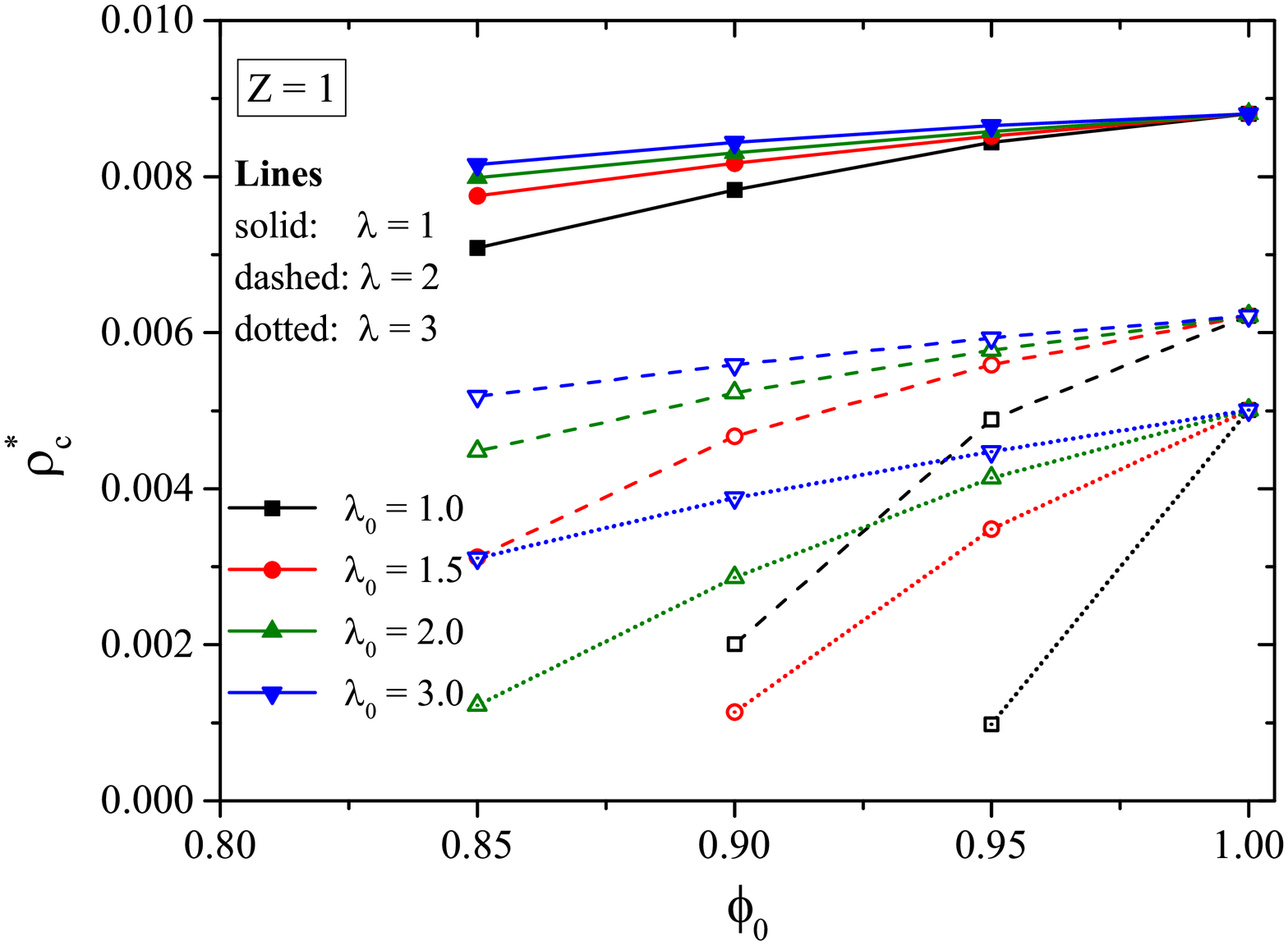}
\caption{\label{fig:critZ1Phi0}
Critical temperature $T^{*}_{c}$ (left panel) and critical density $\rho^{*}_{c}$ (right panel)
of a confined PM fluid as a function of matrix porosity $\phi_{0}$ for $Z=1$.
Solid lines correspond to the case of a PM fluid for $\lambda=1$, dashed lines -- $\lambda=2$
and dotted lines -- $\lambda=3$. Symbols indicate which size of matrix particles $\lambda_{0}$
is used in each case.}
\end{center}
\end{figure}

\begin{figure}[ht]
\begin{center}
\includegraphics[clip,width=0.8\textwidth,angle=0]{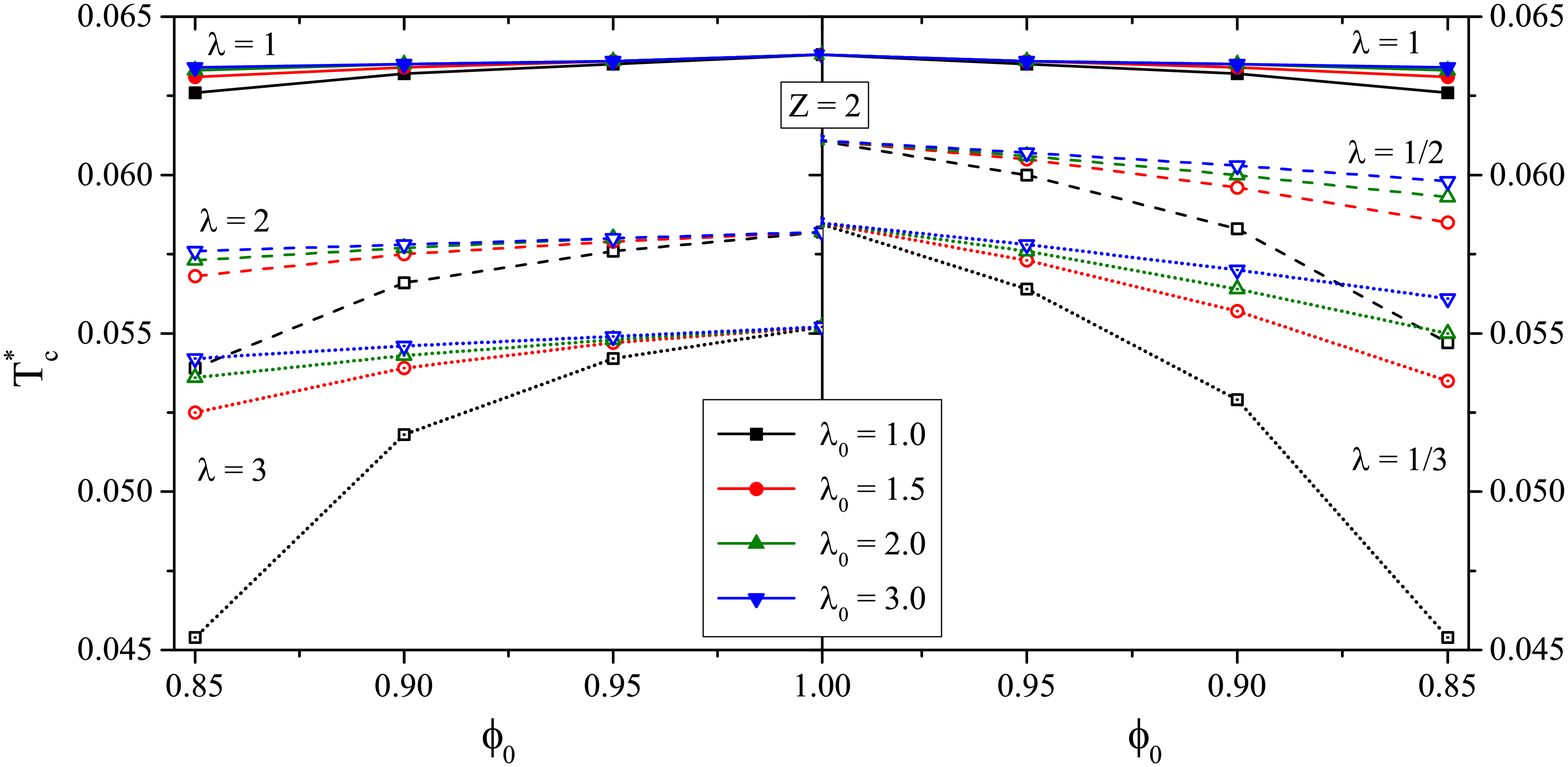}  \\
\includegraphics[clip,width=0.8\textwidth,angle=0]{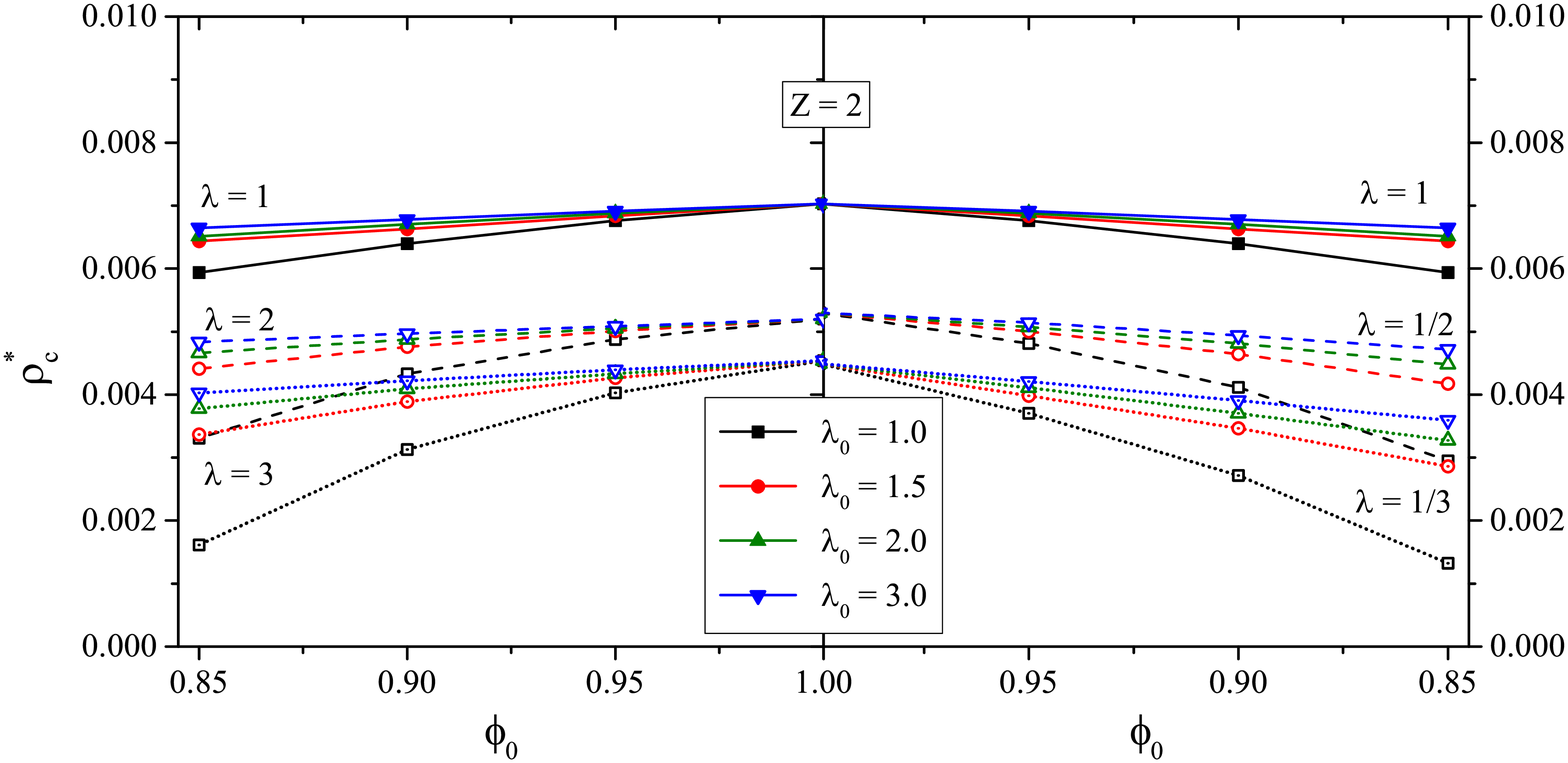}
\caption{\label{fig:critZ2Phi0}
Critical temperature $T^{*}_{c}$ (upper figure) and critical density $\rho^{*}_{c}$ (lower figure)
of a confined PM fluid as a function of matrix porosity $\phi_{0}$ for $Z=2$.
Solid lines correspond to the case of a PM fluid for $\lambda=1$ (left and right panels),
dashed lines -- $\lambda=2$ (left panels) and $\lambda=1/2$ (right panels),
dotted lines -- $\lambda=3$ (left panels) and $\lambda=1/3$ (right panels).
Symbols indicate which size of matrix particles $\lambda_{0}$ is used in each  case.
}
\end{center}
\end{figure}

As it have been mentioned above, a decrease of the porosity leads to the lowering of both the critical temperature and critical density of a confined PM fluid. To analyse the effect of the matrix porosity on the critical parameters the dependencies of $T^{*}_{c}$ and $\rho^{*}_{c}$ on $\phi_{0}$ are presented in Figs.~\ref{fig:critZ1Phi0} and \ref{fig:critZ2Phi0}.
It is found that $T^{*}_{c}$ and $\rho^{*}_{c}$ decrease faster with decreasing the matrix porosity
if the size-asymmetry of ions $\lambda$ is larger. It is also shown that an increase of the size of matrix particle  leads to an increase of both  $T^{*}_{c}$ and $\rho^{*}_{c}$ and the dependency of the critical parameters on $\phi_{0}$ becomes weaker. These trends are similar for the critical temperature and the critical density for $Z=1$ and $Z=2$.
In Fig.~\ref{fig:critZ1Phi0}, because of the symmetry $T^{*}_{c}(\lambda)$=$T^{*}_{c}(1/\lambda)$ and $\rho^{*}_{c}(\lambda)$=$\rho^{*}_{c}(1/\lambda)$ for $Z=1$, we show $T^{*}_{c}(\phi_{0})$ and $\rho^{*}_{c}(\phi_{0})$ only for the size-asymmetry ratio $\lambda\geq1$. By contrast, for $Z=2$ we present the results for the whole range of $\lambda$ considered
in our study. Therefore, it is seen from Fig.~\ref{fig:critZ2Phi0} that  the critical temperatures at
$\lambda<1$ are always higher than the critical temperatures at $\lambda>1$ for the given $\lambda_{0}$ and $\phi_{0}$.
However, the critical density behaves differently. Despite an asymmetry of $\rho^{*}_{c}$ with respect to $\delta=0$ is rather weak,
it is still visible at $\phi_{0}=1$ that $\rho^{*}_{c}$ for $\lambda=2$ is a bit lower than for $\lambda=1/2$, and it is slightly higher for $\lambda=3$ in comparison with $\lambda=1/3$.
These results agree with the results obtained earlier in \cite{Patsahan_Patsahan:11} for a bulk $2:1$ PM fluid.
On the other hand, the matrix presence can make $\rho^{*}_{c}$ lower for $\lambda=1/2$ in comparison with the case of $\lambda=2$.
For instance, it can be clearly observed in a matrix of small particles $\lambda_{0}=1$
and at  low porosity $\phi_{0}=0.85$ (Fig.~\ref{fig:critZ2Phi0}).
It means that for $\lambda<1$ the critical density is more sensitive to the matrix porosity: it decreases faster with
decreasing $\phi_{0}$. At the same time,  this effect is  more pronounced for the critical temperature.

\section{Conclusions}
We have studied the vapour-liquid critical parameters of a charge- and size-asymmetric PM fluid  in a
disordered porous matrix composed of hard-sphere obstacles by using the CVs based theory combined with an extension
of the SPT.  Our calculations are based on an explicit expression for the relevant chemical potential conjugate to the order parameter which
takes into account the third order correlations between ions. In this paper, we have focused  on the particular case  of
ion charge asymmetry, i.e., $Z=q_{+}/q_{-}=2$.

We have calculated the phase diagrams and the corresponding reduced critical temperature $T_{c}^{*}$ and the
reduced critical density $\rho_{c}^{*}$ of a $2$:$1$ asymmetric PM fluid at different values of  size asymmetry ratios
$\lambda=\sigma_{+}/\sigma_{-}$
and $\lambda_{0}=\sigma_{0}/\sigma_{-}$ and at different values of matrix porosity $\phi_{0}$.
Our results have demonstrated  the trends of the critical parameters which are common for  the charge-symmetric and charge-asymmetric
cases: (i) the critical temperature and the critical density are lower when the matrix porosity decreases; (ii) at a fixed porosity,
both  critical parameters are higher in a matrix of large particles than in a matrix of small particles; (iii) an increase in
$\lambda$ leads to the lowering of the critical temperature and the critical density. Despite these general trends we have found
some peculiarities  of the behaviour of the critical parameters of a confined charge-asymmetric ionic fluid. Our main conclusion
is that for $Z=2$ the effect of the matrix presence on the critical parameters is weaker when compared with the charge-symmetric case.

It is worth noting that our analytical theory gives a qualitative picture of the vapour-liquid phase behaviour of a charge- and
size-asymmetric ionic fluid confined in a disordered porous medium. Nevertheless, the  present study, to our best knowledge,
is the first attempt to gain insight into this complicated problem.

\acknowledgments
This project has received funding from the European Unions Horizon 2020 research and
innovation programme under the Marie Sk{\l}odowska-Curie grant agreement No 734276, and from the State Fund For
Fundamental Research (project N F73/26-2017).

\appendix

\section{ Analytical expressions for the chemical potentials of a two-component hard-sphere system confined in a hard-sphere matrix:
the SPT2b approximation}

Recently  \cite{HolovkoDong16}, analytical expressions for
the thermodynamic functions (pressure, Helmholtz free energy, and chemical potentials) of a multicomponent
hard-sphere fluid in a multicomponent matrix have been derived within
the framework of the scaled particle theory  (SPT) extended for the case of confined hard-sphere fluids. Moreover,
in \cite{HolovkoDong16},
the accuracy of various variants obtained
from the basic  SPT formulation is evaluated against the simulation results and it is shown that in most cases the approximation referred
to the SPT2b has the best accuracy.  Based on these results, analytical expressions for the partial chemical potentials of
our reference system consisting of a two-component hard-sphere system confined in a hard-sphere matrix are obtained in the SPT2b
approximation  \cite{PatPatHol17-3}. In particular, the expression for $\nu_{+}^{r}$
is as follows:
\begin{eqnarray}
\nu_{+}^{r}&=&\nu_{+}^{SPT2b}=\ln(\Lambda_{+}^{3}\eta_{+})-\ln(\phi_{+})+k_{+}^{1}\frac{\eta_{i}/\phi_{0}}{1-\eta_{i}/\phi_{0}}
 \nonumber \\
&&
+k_{+}^{2}
\left(\frac{\eta_{i}/\phi_{0}}{1-\eta_{i}/\phi_{0}}\right)^2
+k_{+}^{3}\left(\frac{\eta_{i}/\phi_{0}}{1-\eta_{i}/\phi_{0}}\right)^3-\ln\left(1-\frac{\eta_{i}}{\phi}\right)
 \nonumber \\
&&
\times
\left\{1
-\frac{\phi}{\eta_{i}}\left[1-\frac{\phi}{\phi_{+}}\frac{(\eta_{+}+\lambda^{3}\eta_{-})}{\eta_{i}}\right]\right\}
-\frac{\phi_{0}}{\eta_{i}}\ln\left(1-\frac{\eta_{i}}{\phi_{0}}\right)
\nonumber \\
&&
\times
\left(1-\frac{\eta_{+}
+\lambda^{3}\eta_{-}}{\eta_{i}}\right)
+\frac{(\eta_{+}+\lambda^{3}\eta_{-})}{\eta_{i}}\left(1-\frac{\phi}{\phi_{+}}\right),
\label{nu_+_hs}
\end{eqnarray}
where $\eta_{i}=\eta_{+}+\eta_{-}$ is the packing fraction of the ions, $\eta_{A}=\frac{\pi}{6}\rho_{A}\sigma_{A}^{3}$ ($A=+,-$),
 $\phi_{0}=1-\eta_{0}$ is the geometrical porosity, $\eta_{0}=\pi\rho_{0}\sigma_{0}^{3}/6$,
$\rho_{0}$ is the number density  of the matrix obstacles, and
\begin{equation}
\phi^{-1}=\frac{1}{\eta_{i}}\left(\frac{\eta_{+}}{\phi_{+}}+\frac{\eta_{-}}{\phi_{-}}\right).
\label{phi}
\end{equation}
For  $\phi_{+}$, we have
\begin{eqnarray}
&&\phi_{+}=(1-\eta_{0})\exp\left\{-\frac{6\tau\lambda}{(1+\lambda)}\frac{\eta_{0}}{(1-\eta_{0})}
\left[1+\frac{\tau\lambda}{1+\lambda}
\left(2+\frac{3\eta_{0}}{1-\eta_{0}}\right)
\right.
\right.\nonumber\\
&&
\left.\left.
+\frac{4}{3}\left(\frac{\tau\lambda}{1+\lambda}\right)^{2}
\left(1+\frac{3\eta_{0}}{1-\eta_{0}}
+3\left(\frac{\eta_{0}}
{1-\eta_{0}}\right)^{2}\right)\right]\right\},
\label{phi_+_HS}
\end{eqnarray}
where the parameter
\begin{equation}
\tau=\frac{\sigma_{+-}}{\sigma_{0}}=\frac{1+\lambda}{2\lambda_{0}}
\label{tau}
\end{equation}
is introduced.
$\phi_{-}$ is obtained from (\ref{phi_+_HS}) by replacing $\lambda$ with $1/\lambda$.

The expressions for coefficients $k_{+}^{i}$ are as follows:
\begin{eqnarray}
 k_{+}^{1}&=&\frac{\eta_{-}\lambda(\lambda^{2}+3\lambda+3)+7\eta_{+}}{\eta_{i}}
+\frac{3\tau\lambda\eta_{0}\left[\eta_{-}(\lambda^2+6\lambda+1)+8\eta_{+}\right]}{\eta_{i}(1+\lambda)(1-\eta_{0})}\nonumber \\
&&
+\frac{6\tau^2\lambda^2\eta_{0}(1+2\eta_{0})\left[\eta_{-}(1+\lambda)+2\eta_{+}\right]}{\eta_{i}(1+\lambda)^2(1-\eta_{0})^2},
\label{k+1}
\end{eqnarray}
\begin{eqnarray}
k_{+}^{2}&=&\frac{3}{2}\frac{\left[\eta_{-}^2\lambda^2(2\lambda+3)+2\eta_{+}\eta_{-}\lambda(\lambda+4)+5\eta_{+}^2\right]}{\eta_{i}^2}
\nonumber \\
&&
+\frac{3\tau\lambda\eta_{0}\left[2\eta_{-}^{2}\lambda(2+3\lambda)+\eta_{+}\eta_{-}(\lambda^2+14\lambda+5)+10\eta_{+}^{2}\right]}
{\eta_{i}^2(1+\lambda)(1-\eta_{0})}
\nonumber \\
&&
+\frac{6\tau^2\lambda^2\eta_{0}\left[\eta_{-}(4\eta_{0}\lambda+\eta_{0}+\lambda)+\eta_{+}(5\eta_{0}+1)\right]}{(1+\lambda)^2
(1-\eta_{0})^2\eta_{i}},
\label{k+2}
\end{eqnarray}
\begin{eqnarray}
 k_{+}^{3}=3\left[\frac{2\tau\lambda\eta_{0}}{(1+\lambda)(1-\eta_{0})}+\frac{\eta_{+}
 +\lambda\eta_{-}}{\eta_{i}}\right]^2\frac{\eta_{+}+\lambda
 \eta_{-}}{\eta_{i}}.
 \label{k+3}
\end{eqnarray}
The expression for $\nu_{-}^{r}$ can be obtained from Eqs.~(\ref{nu_+_hs})-(\ref{k+3}) by replacing $\eta_{+}$ with $\eta_{-}$  and vice versa
as well as by replacing $\lambda$ with $1/\lambda$, $\phi_{+}$ with $\phi_{-}$, and $k_{+}^{i}$ with $k_{-}^{i}$ ($i=1,2,3$).


 \end{document}